\documentclass[acmsmall]{acmart}
\usepackage{tikz}

\usepackage{mathtools}
\usepackage{comment}
\usepackage{caption}
\usepackage{url}
\usepackage[T1]{fontenc}
\usepackage{makecell}
\usepackage{multirow}
\usepackage{multicol}
\usepackage{listings}
\usepackage[ruled, linesnumbered, algo2e, noend]{algorithm2e}
\usepackage[noend]{algpseudocode}
\algrenewcommand\algorithmicindent{0.8em}
\usepackage{diagbox}
\usepackage{hhline}
\usepackage{textcomp}
\usepackage{adjustbox}
\usepackage{tabularx}
\newcommand{\expfigwidth}{0.3}
\usepackage{pgf}
\pgfmathsetmacro{\expfigheight}{\expfigwidth * (0.175/0.25)}
\usepackage{graphicx}
\usepackage{subcaption}

\usepackage{balance}
\usepackage{pifont}%
\usepackage{amsmath}
\usepackage{enumitem}
\usepackage{stfloats}
\usepackage{soul}
\usepackage{xcolor}

\newcommand{\squishlist}{
   \begin{list}{$\bullet$}
    { \setlength{\itemsep}{1pt}
      \setlength{\parsep}{0pt}
      \setlength{\topsep}{2pt}
      \setlength{\partopsep}{0pt}
      \setlength{\listparindent}{-2pt}
      \setlength{\itemindent}{-5pt}
      \setlength{\leftmargin}{1.5em}
      \setlength{\labelwidth}{0em}
      \setlength{\labelsep}{0.5em} 
    } 
}
\newcommand{\squishlistend}{
    \end{list}  }
    
\newcommand{\mydashlist}{
   \begin{list}{$-$}
    { \setlength{\itemsep}{1pt}
      \setlength{\parsep}{0pt}
      \setlength{\topsep}{2pt}
      \setlength{\partopsep}{0pt}
      \setlength{\listparindent}{-2pt}
      \setlength{\itemindent}{-5pt}
      \setlength{\leftmargin}{1.5em}
      \setlength{\labelwidth}{0em}
      \setlength{\labelsep}{0.5em} 
    } 
}

\newcommand{\mydashlistend}{
    \end{list}  }

\newcommand{\romanlist}{
   \begin{enumerate}[label=\textit{\roman*)},
    itemsep=2pt,
    parsep=0pt,
    topsep=2pt,
    partopsep=0pt,
    listparindent=0pt,
    itemindent=0pt,
    leftmargin=2em,
    labelsep=0.5em
   ]
}

\newcommand{\romanlistend}{
    \end{enumerate}  }

\newcommand{\revision}[1]{#1}
\newcommand{\hidden}[1]{}

\newcommand{\Blue}[1]{#1}
\newcommand{\Fluorescent}[1]{#1}

\AtBeginDocument{%
  }

\setcopyright{cc}
\setcctype{by}
\acmJournal{PACMMOD}
\acmYear{2026} \acmVolume{4} \acmNumber{6 (SIGMOD)} \acmArticle{325} \acmMonth{12}
\acmDOI{10.1145/3856320}

\begin{document}

\title[Cache Replacement in Direct-Access Cloud OLAP]{Cache Replacement in Direct-Access Cloud OLAP—An Experimental Comparison and Practical Guidance}

\author{Taejun Yoon}
\authornote{Both authors contributed equally to this research.}
\affiliation{%
  \institution{POSTECH}
  \city{Pohang}
  \country{Republic of Korea}}
\email{tjyoon@dblab.postech.ac.kr}

\author{Taesung Lee}
\authornotemark[1]
\affiliation{%
  \institution{POSTECH}
  \city{Pohang}
  \country{Republic of Korea}}
\email{tslee@dblab.postech.ac.kr}

\author{Joo Yeon Lee}
\affiliation{%
  \institution{SAP Labs Korea}
  \city{Seoul}
  \country{Republic of Korea}}
\email{joo.lee@sap.com}

\author{Ki Hong Kim}
\affiliation{%
  \institution{SAP Labs Korea}
  \city{Seoul}
  \country{Republic of Korea}}
\email{ki.kim@sap.com}

\author{Byungchul Tak}
\authornote{Corresponding authors.}
\affiliation{%
  \institution{Kyungpook National University}
  \city{Daegu}
  \country{Republic of Korea}}
\email{bctak@knu.ac.kr}

\author{Wook-Shin Han}
\authornotemark[2]
\affiliation{%
  \institution{POSTECH}
  \city{Pohang}
  \country{Republic of Korea}}
\email{wshan@dblab.postech.ac.kr}

\renewcommand{\shortauthors}{Taejun Yoon et al.}


\received{17 April 2026}
\received[revised]{19 August 2026}
\received[accepted]{12 September 2026}

\newcommand{\REPALGO}[1]{\textsf{#1}}
\newcommand{\LRU}{\REPALGO{LRU}}
\newcommand{\LFU}{\REPALGO{LFU}}
\newcommand{\PerfectLFU}{\REPALGO{Perfect-LFU}}
\newcommand{\Clock}{\REPALGO{Clock}}
\newcommand{\LRUK}{\REPALGO{LRU-K}}
\newcommand{\LRUOne}{\REPALGO{LRU-1}}
\newcommand{\LRUTwo}{\REPALGO{LRU-2}}
\newcommand{\LRUInf}{\REPALGO{LRU-\ensuremath{\infty}}}
\newcommand{\GDS}{\REPALGO{GDS}}
\newcommand{\GDSF}{\REPALGO{GDSF}}
\newcommand{\LRUSP}{\REPALGO{LRU-SP}}
\newcommand{\LFUDA}{\REPALGO{LFU-DA}}
\newcommand{\LIRS}{\REPALGO{LIRS}}
\newcommand{\ARC}{\REPALGO{ARC}}
\newcommand{\SLRU}{\REPALGO{SLRU}}
\newcommand{\SFourLRU}{\REPALGO{S4LRU}}
\newcommand{\Hyperbolic}{\REPALGO{Hyperbolic}}
\newcommand{\FIFO}{\REPALGO{FIFO}}
\newcommand{\SFIFO}{\REPALGO{S3-FIFO}}
\newcommand{\Sieve}{\REPALGO{Sieve}}
\newcommand{\LeCaR}{\REPALGO{LeCaR}}
\newcommand{\LHD}{\REPALGO{LHD}}
\newcommand{\LRB}{\REPALGO{LRB}}
\newcommand{\Cacheus}{\REPALGO{Cacheus}}
\newcommand{\SRLRU}{\REPALGO{SR-LRU}}
\newcommand{\CRLFU}{\REPALGO{CR-LFU}}
\newcommand{\CRPerfectLFU}{\REPALGO{CR-Perfect-LFU}}
\newcommand{\GLCache}{\REPALGO{GL-Cache}}
\newcommand{\ThreeLCache}{\REPALGO{3L-Cache}}
\newcommand{\TwoQ}{\REPALGO{2Q}}
\newcommand{\Crystal}{\REPALGO{Crystal}}
\newcommand{\OURALGO}{\REPALGO{LARA}}
\newcommand{\OURALGOMEANING}{Latency-aware Adaptive Replacement Algorithm}
\newcommand{\OURHR}{Unified Hit Ratio}
\newcommand{\OURHRSHORT}{UHR}

\begin{abstract}
Direct-access cloud OLAP engines rely on local caches to hide remote object-store latency, yet the replacement design space remains poorly understood. Existing systems largely inherit recency- or Clock-oriented policies from DBMS buffer management, even though cloud misses combine a size-independent request overhead with a size-dependent transfer cost and cache objects vary widely in size. We present a broad experimental comparison of 22 cache-replacement algorithms and one semantic-caching baseline across seven analytical workloads and ten cache sizes. To compare them on a common basis, we formulate Unified Hit Ratio (UHR), a latency-weighted hit ratio that decomposes into a workload-specific weighted combination of object and byte hit ratios, and equivalently, the normalized fraction of aggregate remote-download latency saved by caching.
Our results yield practitioner-oriented guidance. Classical size-aware policies developed outside the DBMS literature remain surprisingly strong, dominating object hit ratio on most workloads and staying near the UHR frontier. Byte efficiency depends primarily on aggressive filtering of zero-reuse objects, while under multi-tenant drift lightweight per-request adaptation is often more robust than retraining-heavy learned methods. We instantiate these lessons in LARA, a compact reference policy that validates the guidance through competitive UHR performance.
\end{abstract}

\maketitle

\vspace{-1mm}
\section{Introduction}
\label{sec:intro}

Cloud-native analytical DBMSs such as Snowflake, Amazon Redshift, BigQuery, Databricks SQL, ClickHouse, and MotherDuck increasingly operate over compute-storage-disaggregated architectures, in which data resides in remote object stores and compute nodes access it over the network~\cite{snowflake,redshift,bigquery,databricks,clickhouse,motherduck,dong2024cloud,amazons3,azureblobstorage,googlecloudstorage}.
This architecture improves elasticity, scalability, and cost efficiency, but it also makes query execution sensitive to remote-object-store latency and bandwidth, which in turn makes caching near compute a first-order performance mechanism~\cite{dong2024cloud,durner2023exploiting,vuppalapati2020building,alluxio}.
By \emph{direct-access} cloud OLAP, we mean engines that fetch raw remote files or chunks into a local cache and process them locally, rather than caching predicate-defined semantic regions or relying on remote predicate pushdown.
This makes it hard to answer a simple operational question: \emph{which cache-replacement family should a direct-access cloud OLAP engine deploy, under which workload conditions?}

While the deployment model is new, its local cache-management choices are largely inherited from classical DBMS buffer management~\cite{lruk,2q,clock,vmcache,neumann2020umbra}.
Publicly described systems illustrate this pattern: Amazon Redshift describes a two-level Clock-based local caching scheme in its managed-storage path~\cite{redshift}; Databricks documents recency-based local-file caching~\cite{databricksdiskcache}; DuckDB exposes recency-based caching for external files~\cite{duckdb_external_file_cache}; and Snowflake emphasizes extensive local-data caching in its elastic architecture~\cite{snowflake,vuppalapati2020building}.
This inheritance is natural, because classical buffer managers were designed for page-oriented caches and favored simple, low-overhead, and operationally robust heuristics~\cite{lruk,2q,clock}.

However, the assumptions behind page-oriented buffering do not carry over cleanly to cloud OLAP.
Analytical engines over Parquet-like data often read at file, row-group, or column-chunk granularity~\cite{parquet_concepts,open_formats,hana}, which implies that cacheable objects can vary widely in size.
A miss to a remote object store combines a size-independent request overhead, such as network round-trip time (RTT) and request setup, with a size-dependent transfer cost~\cite{durner2023exploiting}.
Moreover, production cloud-analytics workloads exhibit multi-tenancy, burstiness, repeated subworkloads, and even DML-driven evolution~\cite{cab,redset,redbench,yin2025mtd}.
Once these properties arise together, replacement policies designed under page-oriented, roughly uniform-cost assumptions do not necessarily optimize the quantities that matter most in cloud storage.

These changed assumptions make ideas from adjacent caching domains newly relevant.
Web and storage-system caching have long studied variable-sized objects and heterogeneous miss costs via size-aware policies such as \GDS{}, \GDSF{}, \LRUSP{}, and \LFUDA{}~\cite{gds,gdsf,lrusp,lfuda}; scan resistance and one-hit-wonder filtering via \LIRS{}, \SFIFO{}, and \Sieve{}~\cite{lirs,s3fifo,sieve}; and lightweight online adaptivity via \ARC{}, \LeCaR{}, and \Cacheus{}~\cite{arc,lecar,cacheus}.
Yet, to the best of our knowledge, these policy families have not been compared side by side under a common direct-access cloud OLAP setting, nor distilled into practical guidance for system builders deciding which replacement family to deploy.

Doing so is nontrivial because prior work differs in both system model and evaluation target.
Some systems cache physical objects, whereas others cache semantic regions defined by query predicates~\cite{dar1996semantic,ren2003semantic,durner2021crystal}.
Broadly, some systems study replacement for direct access to raw files~\cite{durner2023exploiting,hana}, whereas others solve a larger problem of caching with remote predicate evaluation or pushdown-aware execution~\cite{yang2021flexpushdowndb,yang2024flexpushdowndb,s3select}.
Even the evaluation target is unsettled: Object Hit Ratio (OHR) captures miss count, Byte Hit Ratio (BHR) captures missed volume, and delay-oriented metrics such as DSR were developed for proxy caches with different delay models~\cite{dsr}.
A meaningful cross-family comparison requires a common experimental basis and a latency-oriented evaluation metric.

The primary contribution of this paper is therefore a clean-slate experimental comparison and practitioner-oriented selection guidance, not a claim that one newly proposed policy universally dominates.
We focus on direct-access cloud OLAP engines that cache \emph{physical column chunks}, i.e., contiguous per-column storage units within a row group in columnar formats such as Parquet~\cite{parquet_concepts,hana}, while fetching raw data from a remote object store.
Within this scope, we port 22 representative replacement algorithms from DBMS buffer management, web caching, and storage-system caching into a common framework, enforce the same cache unit and pin/unpin semantics across all policies, and compare them against a semantic-caching baseline.
Our goal is to answer a practical question: \emph{which policy families work best under which workload characteristics, why do they do so, and what trade-offs should practitioners expect?}

To make such guidance actionable, we establish both a common experimental basis and a latency-oriented evaluation metric.
Because OHR and BHR each capture only part of remote-I/O benefit, we formulate \OURHR{} (\OURHRSHORT{}), a latency-weighted hit ratio under our cloud miss-cost model.
\OURHRSHORT{} can be written as a workload-specific weighted combination of OHR and BHR and, equivalently, as the normalized fraction of aggregate remote-download latency saved by caching; Section~\ref{sec:metric-lens} derives this equivalence explicitly.
We validate \OURHRSHORT{} against live S3 replay and then use OHR, BHR, and \OURHRSHORT{} together not only for black-box ranking, but also for white-box explanation of which workload traits favor which policy families.

Our results tell a more specific story than the generic statement that ``no single policy wins everywhere.''
Classical size-aware policies originally developed outside the DBMS literature---notably \GDS{}, \GDSF{}, and \LRUSP{}---dominate OHR on most workloads and remain near the \OURHRSHORT{} frontier.
Strong BHR, in contrast, comes primarily from aggressive filtering of zero-reuse objects, as seen in \Cacheus{}, \SFIFO{}, and \LIRS{}.
Under multi-tenant drift, lightweight per-request adaptivity is often more robust than retraining-heavy learned methods.
These findings yield actionable selection guidance: object size is the dominant factor for OHR, zero-reuse filtering is critical for byte efficiency under tight caches, and online adaptivity matters when workloads drift.

Finally, to show that these empirical lessons are constructive rather than merely descriptive, we instantiate them in a compact reference policy, \OURALGO{} (\emph{\OURALGOMEANING{}}).
\OURALGO{} should be read as a reference instantiation of the empirical guidance rather than the paper's primary contribution; nevertheless, it validates the guidance by achieving the best mean \OURHRSHORT{} rank in our evaluation, while an OHR-tuned variant achieves the best mean OHR rank.

In summary, we make the following contributions:
\squishlist
    \item We establish, to our knowledge, the first broad side-by-side experimental comparison of 22 cache-replacement algorithms plus a semantic-caching baseline for direct-access cloud OLAP under compute-storage disaggregation.
    \item We formulate \OURHR{} (\OURHRSHORT{}), a latency-oriented metric that exposes the workload-dependent balance between request overhead and byte transfer, and validate it against live object-store replay.
    \item We derive practical selection guidance on when size-awareness, zero-reuse filtering, online adaptivity, and semantic caching matter.
    \item We validate that these empirical lessons are constructive with \OURALGO{}, a compact reference instantiation that achieves the best mean \OURHRSHORT{} rank in our evaluation.
\squishlistend

\vspace{-1mm}
\section{Background and Related Work}
\label{sec:background}

This section clarifies what is compared in our study.
We first review the standard metrics used to quantify cache benefit and explain why direct-access cloud OLAP needs a latency-oriented view in addition to traditional hit ratios.
We then position our scope within cloud OLAP caching by distinguishing physical chunk caching from semantic and pushdown-aware alternatives.
Finally, we explain why a meaningful comparison must include algorithms from DBMS buffer management, web caching, and storage-system caching.

\vspace{-2mm}
\subsection{Metrics for Cache Benefit}
\label{subsec:metrics}

A cache study is only useful if its metric reflects the cost that the cache is meant to reduce.
The two standard metrics are \emph{Object Hit Ratio} (OHR) and \emph{Byte Hit Ratio} (BHR), which capture request-level and byte-level efficiency, respectively.
OHR asks how often remote requests are avoided, whereas BHR asks how many remote bytes are avoided.
\Blue{
This distinction is crucial under byte-limited cache capacity.
Because OHR counts every hit equally, retaining several small reusable chunks creates more opportunities for object hits than spending the same capacity on a few large chunks.
Size-aware policies can therefore have a structural advantage under OHR.}
Both remain informative throughout this paper, but direct-access cloud OLAP also requires a latency-oriented view because a miss combines fixed request overhead with size-dependent transfer cost.
We therefore report OHR and BHR throughout and introduce a cloud-specific latency-oriented metric in Section~\ref{sec:metric-lens}.

\vspace{-2mm}
\subsection{Caching Designs in Cloud OLAP}
\label{subsec:cloud-olap-caching}

Cloud OLAP caching is not a single design point.
The most relevant distinction for our study is between \emph{direct-access physical caching}, \emph{semantic caching}, and \emph{pushdown-aware caching}.
Direct-access physical caching keeps raw data objects close to compute and serves requests locally on a hit.
In modern file-based analytical engines, the natural cache units are physical storage objects such as files, row groups, or column chunks rather than fixed-size pages~\cite{parquet_concepts,open_formats,hana}.
This is the architecture most directly associated with local SSD/NVMe caches in direct-access engines~\cite{redshift,databricksdiskcache,duckdb_external_file_cache}.

Semantic caching stores logical regions defined by query semantics rather than physical storage units.
\Crystal{} is a representative recent system in this design space: it caches predicate-defined regions and exploits overlap across analytical queries~\cite{durner2021crystal}.
Because semantic caching changes both the cache unit and the replacement problem, it is not directly comparable to physical chunk caching on a per-policy basis.
We nevertheless include \Crystal{} as a baseline because it is the most relevant cloud-OLAP system that exploits analytical predicate reuse over remote object storage.

Pushdown-aware caching jointly reasons about local caching and remote predicate evaluation.
FlexPushdownDB is the canonical example: it studies when data should be fetched in and when predicate evaluation should instead be pushed into object storage~\cite{yang2021flexpushdowndb,yang2024flexpushdowndb,s3select}.
Similarly, LBSC, \emph{a cost-aware caching framework for cloud databases}, studies heterogeneous refetch penalties~\cite{ji2024lbsc}.
These systems solve a broader problem than replacement alone because they couple caching with remote execution choices.

This paper deliberately narrows the scope to \emph{direct-access caching of physical column chunks}.
That scope matches common file-based analytical execution over open columnar formats~\cite{open_formats,hana}, keeps the cache unit fixed across algorithms, and provides a clean basis for comparing replacement policies without conflating them with query-level transformations such as rewriting a new request against already cached predicate-defined regions, or with pushdown planning such as deciding whether a predicate should be evaluated remotely.\looseness=-1

\vspace{-3mm}
\subsection{Why Policies from Multiple Domains Matter}
\label{subsec:why-multi-domain}

The policy space relevant to direct-access cloud OLAP is broader than current DBMS practice.
The mapping from cloud-OLAP traits to prior algorithmic domains is straightforward: variable object sizes and heterogeneous miss costs motivate web-caching policies; scan-heavy, bursty, and multi-tenant streams motivate storage-system policies; and current operational practice plus low-overhead reuse verification motivate DBMS buffer-management policies.
Current systems naturally inherit simple recency- or Clock-oriented policies from classical DBMS buffer management because those policies are low-overhead and operationally robust~\cite{lruk,2q,clock,vmcache,neumann2020umbra}.
However, once cloud misses combine fixed request overhead with size-dependent transfer cost and cache objects vary substantially in size, the design assumptions of page-oriented buffering become less adequate.

Web caching is relevant because it has long studied variable-sized objects and heterogeneous miss costs.
This produced size- and cost-aware policies such as \GDS{}, \GDSF{}, \LFUDA{}, and \LRUSP{}~\cite{gds,gdsf,lfuda,lrusp}, which explicitly trade off reuse against object size or fetch cost.
Those become directly relevant when cloud OLAP replacement must balance request avoidance against byte avoidance.

Storage-system caching is relevant because cloud OLAP workloads contain scans, one-hit objects, bursty shifts, and multi-tenant interference.
This line produced scan-resistant and reuse-filtering mechanisms such as \LIRS{}, \SFIFO{}, and \Sieve{}~\cite{lirs,s3fifo,sieve}, as well as adaptive designs such as \ARC{}, \LeCaR{}, and \Cacheus{}~\cite{arc, lecar,cacheus}.
These techniques matter because many cloud-analytics traces contain exactly the patterns that punish plain recency or plain frequency policies.

For these reasons, our benchmark includes representative algorithms from DBMS buffer management, web caching, and storage-system caching.
The point of this breadth is not to crown a single universally best method, but to establish which policy families are appropriate under which workload traits.

\vspace{-2mm}
\section{Cache Replacement Algorithms}
\label{sec:replacement-algos}

This section defines the policy space evaluated in our benchmark and introduces the vocabulary used to explain the results.
We group the 22 policies by the primary signal driving eviction.
Section~3.2 covers recency-, frequency-, and queue-based low-overhead baselines, including reuse-filtering variants.
Section~3.3 adds object size and miss cost.
Section~3.4 adds online adaptation and learned prediction.
The goal is not to survey the history of caching, but to make the later OHR, BHR, and UHR analyses easy to interpret.
We use \emph{policy} and \emph{cache replacement algorithm} interchangeably.
Although the original papers use different cache units (pages, files, blocks, or web objects), in our experiments every policy is applied to the same cache unit: a physical column chunk.

\vspace{-2mm}
\subsection{Taxonomy and Terminology}

We classify each policy by four basic signals:
\squishlist
\item \textbf{Recency (R):} how recently an object was accessed.
\item \textbf{Frequency (F):} how many times an object was accessed.
\item \textbf{Insertion order (I):} the order in which objects entered the cache.
\item \textbf{Size (S):} the storage size of the object.
\squishlistend

A signal is \emph{primary} if it directly affects which object is evicted.
Signals used only to break ties are ignored in the class label.
For example, an LFU implementation may break equal-frequency ties by recency, but we still classify it as F-type because frequency remains the main ranking signal.
In principle, 16 combinations of R, F, I, and S are possible.
In practice, only 8 appear in our benchmark: size never appears alone, and no representative policy in our set uses insertion order together with size, or as part of a three-way combination.
Insertion order is fixed at admission, whereas recency is updated on a hit.
Accordingly, \Clock{} and \Sieve{} are RI-type: eviction scans objects in queue order, but a one-bit \emph{reference flag} set on a hit changes how a hit object is treated.

Several terms recur throughout this section.
A \emph{scan} is a long sequence of requests that touches many objects once or very few times.
A policy is \emph{scan-resistant} if such a sequence does not flush out frequently reused objects.
A \emph{one-hit object} is an object that is admitted to the cache and never hit again before eviction.
A \emph{ghost list} or \emph{ghost queue} stores only lightweight metadata of recently evicted objects, not the data itself.
A \emph{probationary queue} is a temporary queue for newly admitted objects that must prove reuse before receiving stronger protection.
A \emph{segmented} policy maintains multiple internal queues or regions and moves objects among them according to their reuse status.

\begin{figure*}[t]
    \centering
    \begin{subfigure}[t]{0.245\textwidth}
        \includegraphics[width=\textwidth]{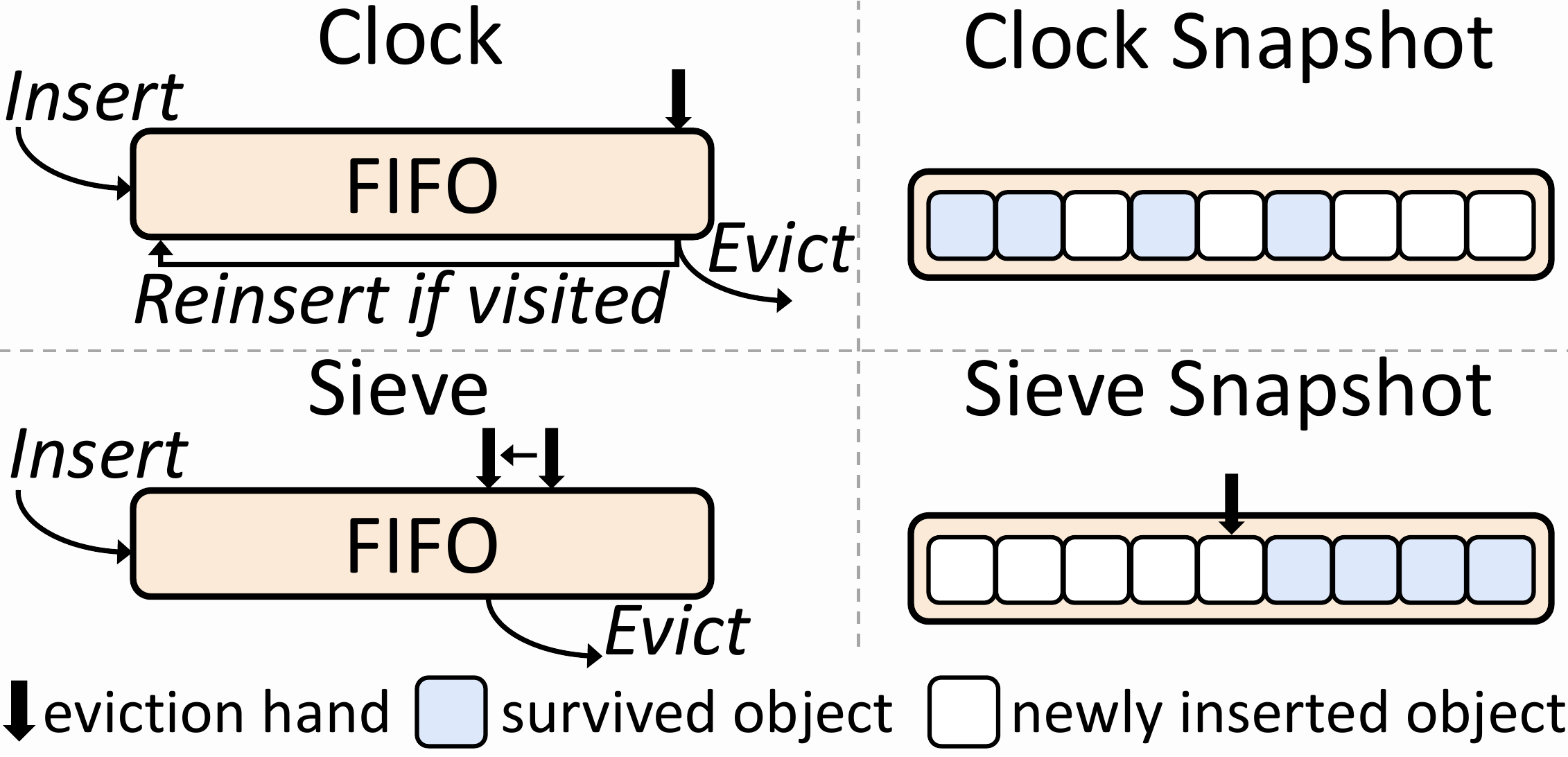}
        \caption{\Clock{} and \Sieve{}}
        \label{fig:clock_sieve}
    \end{subfigure}
    \hfill
    \begin{subfigure}[t]{0.245\textwidth}
        \includegraphics[width=\textwidth]{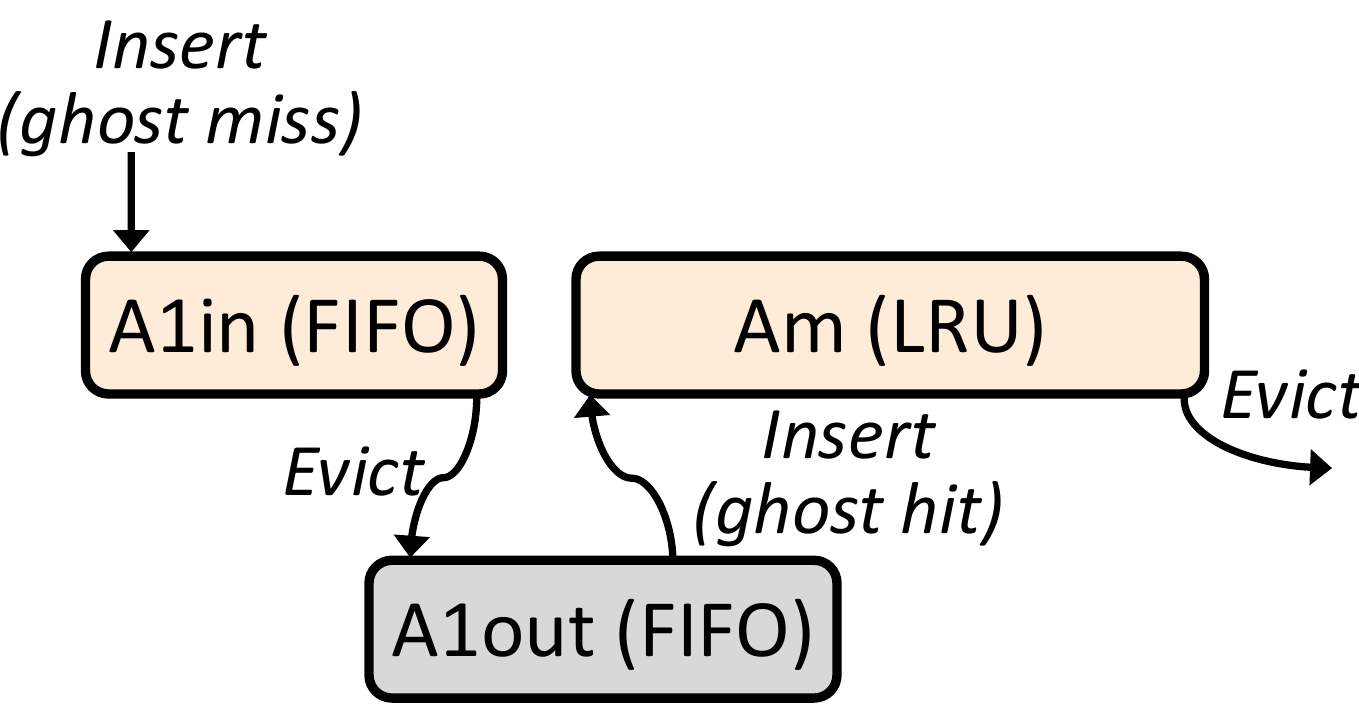}
        \caption{\TwoQ{}}
        \label{fig:2q}
    \end{subfigure}
    \hfill
    \begin{subfigure}[t]{0.245\textwidth}
        \includegraphics[width=\textwidth]{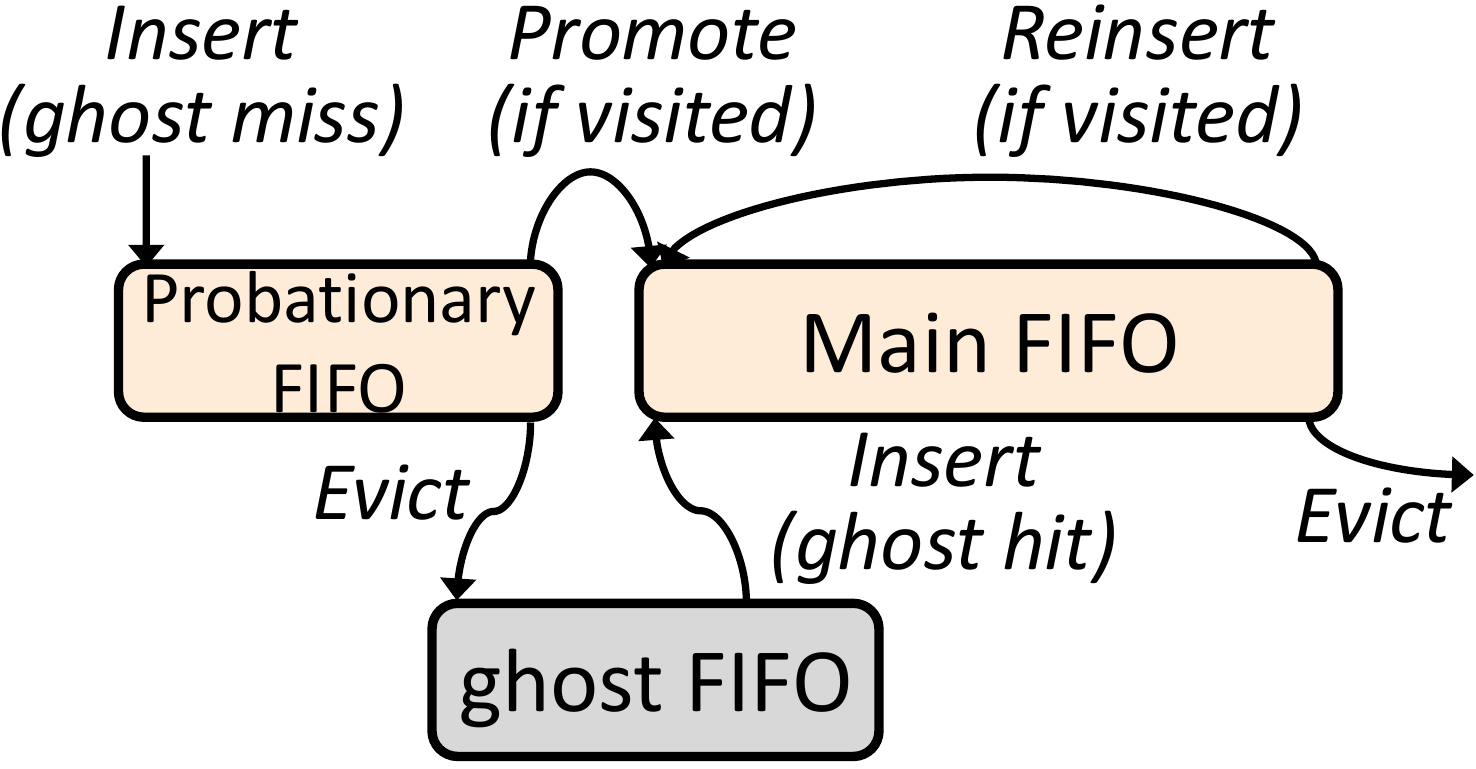}
        \caption{\SFIFO{}}
        \label{fig:s3fifo}
    \end{subfigure}
    \hfill
        \begin{subfigure}[t]{0.245\textwidth}
        \includegraphics[width=\textwidth]{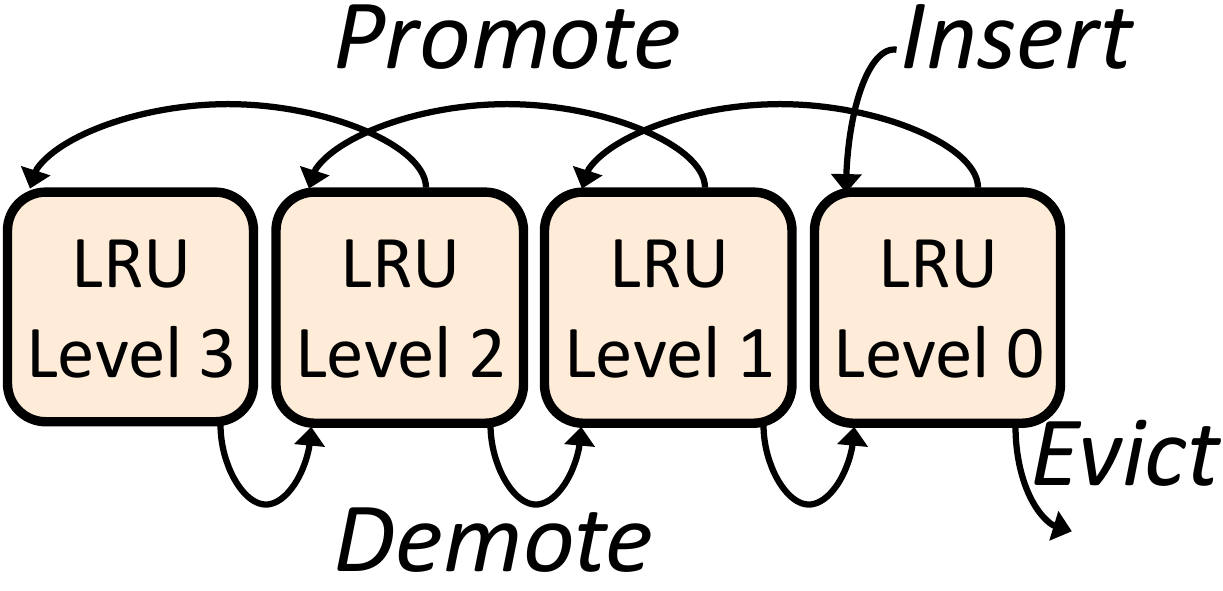}
        \caption{\SFourLRU{}}
        \label{fig:s4lru}
    \end{subfigure}
    \vspace{-10pt}
    \caption{Representative queue-based, ghost-based, and segmented mechanisms. Clock reinserts objects after a second chance, whereas Sieve leaves them in place. 2Q verifies reuse before promotion into the main cache. S3-FIFO uses probationary, main, and ghost queues to filter one-hit objects. S4LRU promotes objects upward on hits and demotes them downward on overflow.}
    \label{fig:algorithm_illustration_clock_sieve_s3fifo_2q_arc}
    \vspace{-8pt}
\end{figure*}

\begin{figure*}[t]
    \centering
    \begin{subfigure}[t]{0.225\textwidth}
        \includegraphics[width=\textwidth]{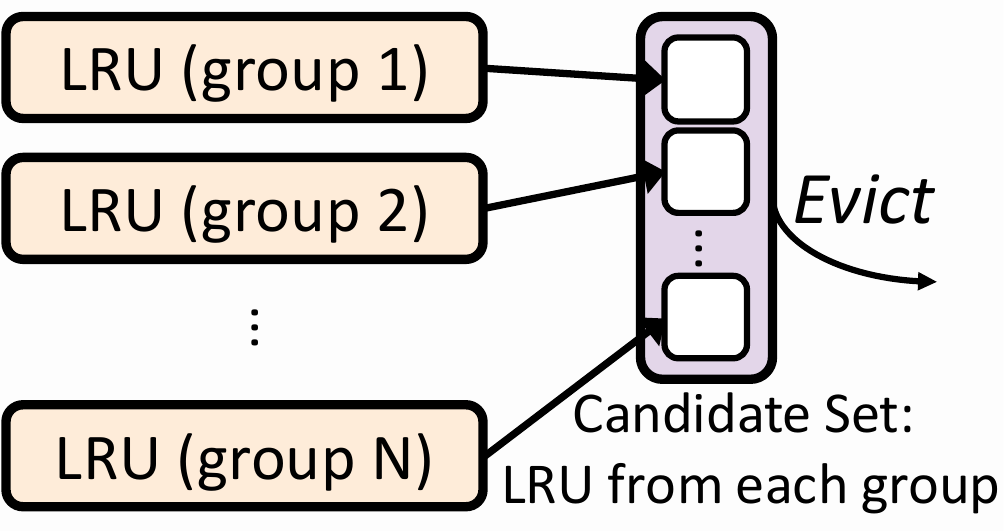}
        \caption{\LRUSP{}}
        \label{fig:lrusp}
    \end{subfigure}
    \hfill
    \begin{subfigure}[t]{0.225\textwidth}
        \includegraphics[width=\textwidth]{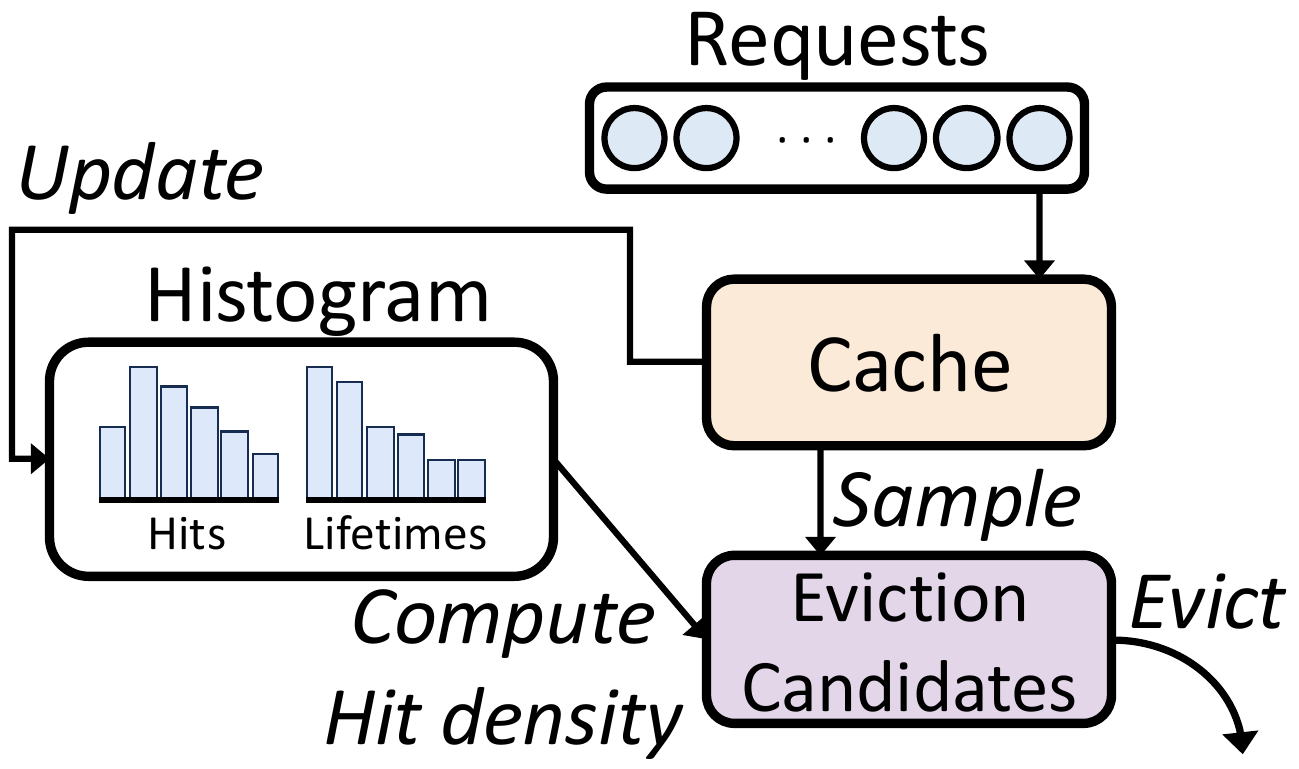}
        \caption{\LHD{}}
        \label{fig:lhd}
    \end{subfigure}
    \hfill
    \begin{subfigure}[t]{0.245\textwidth}
        \includegraphics[width=\textwidth]{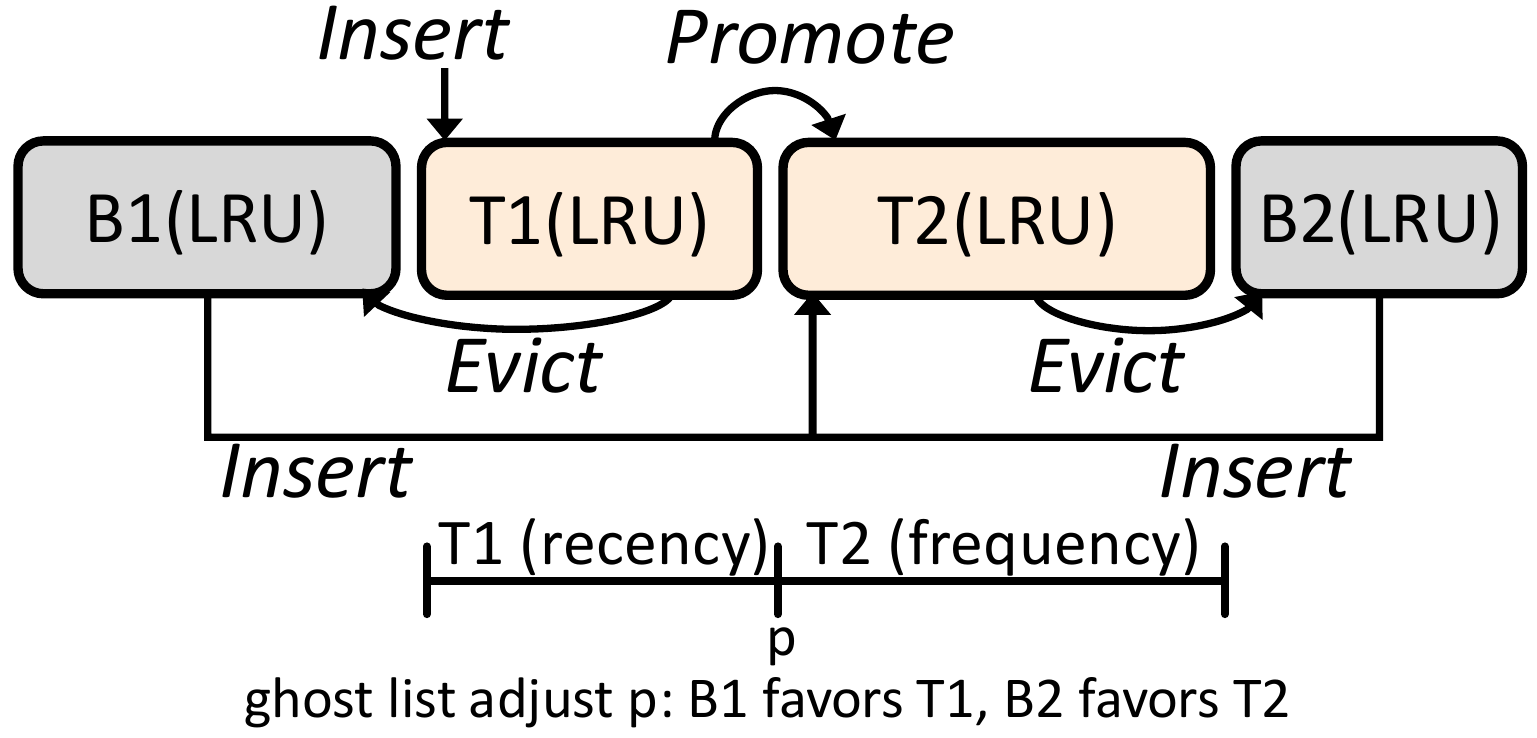}
        \caption{\ARC{}}
        \label{fig:arc}
    \end{subfigure}
    \hfill
    \begin{subfigure}[t]{0.225\textwidth}
        \includegraphics[width=\textwidth]{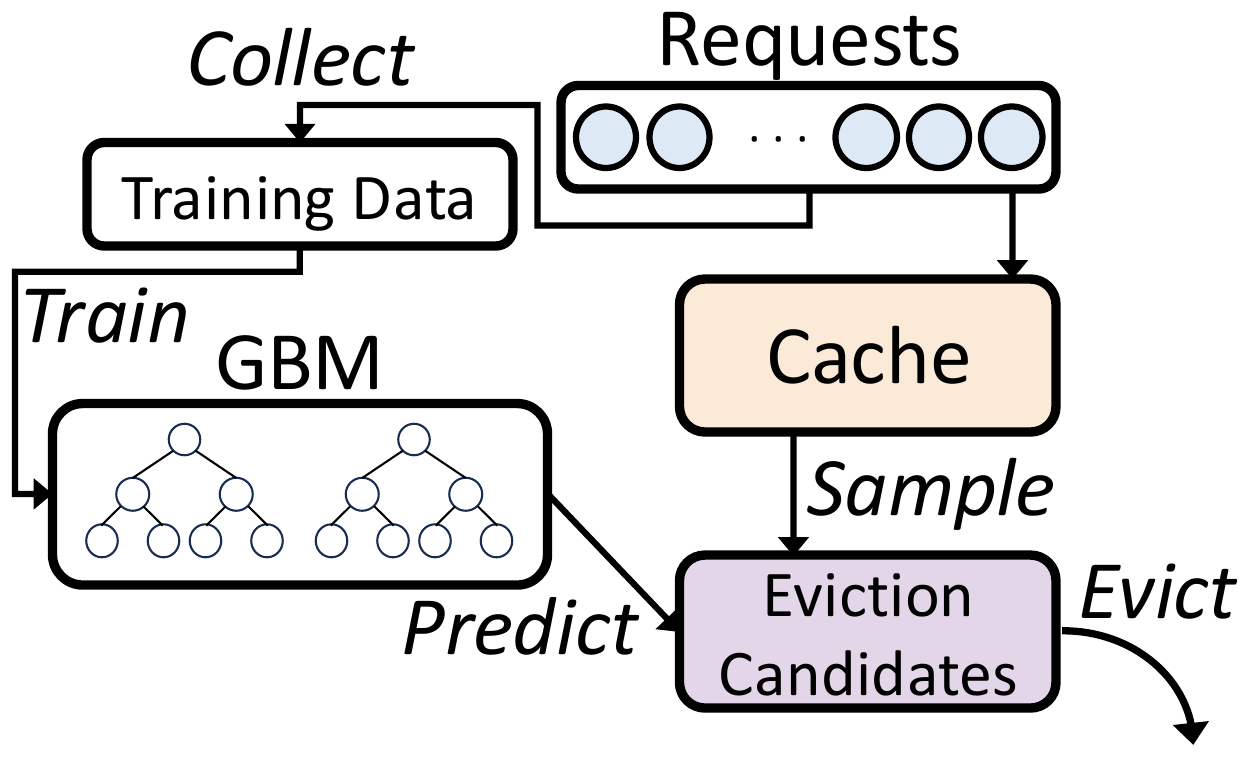}
        \caption{\LRB{}}
        \label{fig:lrb}
    \end{subfigure}
    \vspace{-10pt}
    \caption{Representative size-aware, adaptive, and learned policies. LRU-SP compares queue-tail candidates across size-adjusted groups. LHD estimates hit density from class-specific histograms. ARC adapts the split between recency-oriented and frequency-oriented regions using ghost feedback. LRB samples candidates and predicts their next use with a gradient boosting machine.}
    \label{fig:algorithm_illustration_s4lru_lhd_lrusp_lrb}
    \vspace{-12pt}
\end{figure*}

Table~\ref{tab:cache-algo-comparison} lists all evaluated policies with their class and four
practical attributes (Learning, Domain, Adaptivity,
Segmented). Figures~\ref{fig:algorithm_illustration_clock_sieve_s3fifo_2q_arc}--\ref{fig:algorithm_illustration_s4lru_lhd_lrusp_lrb} illustrate representative mechanisms.

\begin{table}[t]
\centering
\footnotesize
\setlength{\tabcolsep}{4pt}
\renewcommand{\arraystretch}{1.1}
\caption{Cache replacement algorithms evaluated.}
\label{tab:cache-algo-comparison}
\vspace{-10pt}

\begin{tabular}{|l|l|c|l|c|c|}
\hline
\textbf{Algorithm} &
\textbf{class} &
\textbf{Learning} &
\textbf{Domain} &
\textbf{Adaptivity} &
\textbf{Segmented} \\
\hline
\hline
\LRU{}                              & R   & No  & Generic   & Low   & No  \\
\hline
\LFU{}                              & F   & No  & Generic   & Low   & No  \\
\hline
\FIFO{}                             & I   & No  & Generic   & Low   & No  \\
\hline
\Clock{}~\cite{clock}               & RI  & No  & OS        & Low   & No  \\
\Sieve{}~\cite{sieve}               & RI  & No  & Web       & Low   & No  \\
\hline
\SFIFO{}~\cite{s3fifo}              & FI  & No  & Storage   & Low   & Yes \\
\hline
\GDS{}~\cite{gds}                   & RS  & No  & Web       & Low   & No  \\
\hline
\LRUK{}~\cite{lruk}                 & RF  & No  & DBMS      & Low   & No  \\
\TwoQ{}~\cite{2q}                   & RF  & No  & DBMS      & Low   & Yes \\
\LFUDA{}~\cite{lfuda}               & RF  & No  & Web       & Low   & No  \\
\LIRS{}~\cite{lirs}                 & RF  & No  & Storage   & Low   & Yes \\
\ARC{}~\cite{arc}                   & RF  & No  & Storage   & High  & Yes \\
\SFourLRU{}~\cite{s4lru}            & RF  & No  & Web       & Low   & Yes \\
\LeCaR{}~\cite{lecar}               & RF  & Yes & Storage   & High  & No  \\
\Cacheus{}~\cite{cacheus}           & RF  & Yes & Storage   & High  & Yes  \\
\Hyperbolic{}~\cite{hyperbolic}     & RF  & No  & Web       & Low   & No  \\
\hline
\LHD{}~\cite{lhd}                   & RFS & No  & Web       & Medium & No  \\
\GDSF{}~\cite{gdsf}                 & RFS & No  & Web       & Low   & No  \\
\LRUSP{}~\cite{lrusp}               & RFS & No  & Web       & Low   & No  \\
\LRB{}~\cite{lrb}                   & RFS & Yes & Web       & Medium & No  \\
\GLCache{}~\cite{glcache}           & RFS & Yes & Storage   & Medium & No  \\
\ThreeLCache{}~\cite{3lcache}       & RFS & Yes & Web       & Medium & No  \\
\hline
\end{tabular}
\vspace{-8pt}
\end{table}

\vspace{-2mm}
\subsection{Low-Overhead and Reuse-Filtering Policies}

These policies are closest to inherited DBMS practice or low-overhead systems design, and they expose the baseline trade-offs among recency, frequency, insertion order, scan resistance, and explicit reuse verification.

\textit{LRU, LFU, and FIFO.}
\LRU{} (R) reacts fastest to working-set shifts but can evict temporarily cold reusable objects.
\LFU{} (F) preserves globally popular objects but stale counts can remain after phase changes.
\FIFO{} (I) is the simplest queue baseline, yet it cannot distinguish reusable objects from one-hit objects.

\textit{\Clock{} and \Sieve{}.}
\Clock{} and \Sieve{} (Figure~\ref{fig:clock_sieve}) both start from FIFO-style queue order and add a one-bit reference flag that is set on a hit.
In \Clock{}, an eviction hand moves \emph{circularly} through the queue until it finds an object whose flag is clear; if the flag is set, the object receives a \emph{second chance}, the flag is cleared, and the object is reinserted at the head of the queue.
This makes \Clock{} a low-overhead approximation of \LRU{}.
\Sieve{} uses the same flag, but it does not reinsert surviving objects.
Newly admitted objects therefore stay near the head, while long-lived reusable objects drift away from the eviction point, which makes \Sieve{} more aggressive at removing newly admitted but never-reused objects.

\textit{\LRUK{} and \TwoQ{}.}
A scan-resistant policy should prevent a long stream of one-time accesses from flushing out frequently reused objects.
\LRUK{} does this by remembering the last $K$ access times of each object and evicting the object whose $K$-th most recent access lies farthest in the past.
Objects seen fewer than $K$ times are therefore the first candidates for eviction; we evaluate the standard choice $K=2$.
\TwoQ{} achieves a similar effect with lower overhead (Figure~\ref{fig:2q}).
New objects first enter a small \FIFO{} queue, \emph{A1in}.
When they leave this queue, only their identifiers remain in a ghost queue, \emph{A1out}.
An object is promoted to the main \LRU{} queue, \emph{Am}, only if it is referenced again while its identifier is still in \emph{A1out}.

\textit{\LIRS{}, \SFIFO{}, \SFourLRU{}, and \Hyperbolic{}.}
\LIRS{} uses \emph{inter-reference recency} (IRR), defined as the distance in the request stream between the last two references to the same object.
Objects with small IRR are kept in a large protected \emph{LIR} region, whereas objects with large IRR stay in a small \emph{HIR} region from which eviction occurs.
New objects start as HIR.
If a HIR object is referenced again and its updated IRR becomes smaller than that of the oldest LIR object, it is promoted to LIR and the oldest LIR object is demoted to HIR.
\SFIFO{} (Figure~\ref{fig:s3fifo}) uses a probationary queue, a main queue, and a ghost queue.
New objects that leave the probationary queue without a re-reference are filtered out, while promptly re-referenced objects are promoted into the main queue.
\SFourLRU{} (Figure~\ref{fig:s4lru}) extends segmented \LRU{} to four protection levels.
New objects enter the lowest level, hits promote them upward by one level, overflow demotes objects downward, and eviction ultimately occurs from the lowest level.
\Hyperbolic{} avoids maintaining a strict global order over the whole cache.
Instead, it samples a small candidate set and evicts the weakest sampled object according to a cheap score such as
\[
\mathit{Pr}(o)=\frac{n_o}{t_o},
\]
where $n_o$ is the access count and $t_o$ is the number of requests since insertion.
Sampling avoids the cost of a global priority queue.

\vspace{-2mm}
\subsection{Size- and Cost-Aware Policies}
\label{subsec:size-and-cost-aware-polices}

This group adds object size or miss cost to the eviction score.
These policies are central to our study because direct-access cloud OLAP departs from page-like buffering precisely through variable-sized objects and heterogeneous miss penalties.

\textit{\GDS{}, \LFUDA{}, and \GDSF{}.}
GreedyDual-Size (\GDS{}) combines recency, miss cost, and size in a single priority:
\[
  \mathit{Pr}(o) = \mathit{L} + \frac{\mathit{Cost}(o)}{S(o)}
\]
where $S(o)$ is object size, $\mathit{Cost}(o)$ is the miss cost assigned to object $o$, and $\mathit{L}$ is an \emph{inflation value} updated to the priority of the most recently evicted object.
The object with the smallest priority is evicted.
With $\mathit{Cost}(o)=1$, \GDS{} optimizes for OHR, while alternative cost functions can target latency or network traffic reduction.
\LFUDA{} extends \LFU{} with dynamic aging:
\[
   \mathit{Pr}(o) = \mathit{L} + \mathit{Freq}(o)\times \mathit{Cost}(o)
\]
where $\mathit{L}$ is updated exactly as in \GDS{}, namely to the priority of the most recently evicted object.
This prevents objects that were popular in the distant past from remaining protected forever.
\GDSF{} combines the two ideas:
\[
  \mathit{Pr}(o) = \mathit{L} + \frac{\mathit{Freq}(o)\times \mathit{Cost}(o)}{S(o)}
\]
Here again, $\mathit{L}$ follows the same update rule as in \GDS{}.
A compact way to read this is \emph{GDSF = GDS + frequency in the numerator}: relative to \GDS{}, it additionally rewards repeated reuse, and relative to \LFUDA{}, it makes the score explicitly size-aware.

\subsubsection{\revision{Standard and latency-cost variants.}}
\label{sec:latencycost}
\Blue{Section~\ref{subsec:latency-model} defines the modeled remote-download latency of object $o$ in Eq.~\eqref{eq:cloud-latency} as $L(o)=\alpha+\beta S(o)$, where $\alpha$ is the size-independent per-request latency, $\beta$ is the transfer latency per byte, and $S(o)$ is the object size.
For \GDS{}, \GDSF{}, and \LFUDA{}, we evaluate both their standard form and latency-cost variants.
In the standard form, $\mathit{Cost}(o)=1$ and every miss receives the same cost.
In the latency-cost variants, $\mathit{Cost}(o)=L(o)$, which makes the eviction priority reflect this modeled latency.
This comparison tests whether setting $\mathit{Cost}(o)=L(o)$ is sufficient to reduce remote-download latency or not.}

\subsubsection{\LRUSP{} and \LHD{}}
\label{sec:lrusplhd}
\LRUSP{} (Figure~\ref{fig:lrusp}) approximates a size- and popularity-aware \LRU{} by grouping objects with similar $S(o)/\mathit{Freq}(o)$ values into separate LRU queues and then comparing only queue-tail candidates across groups.
This lets it combine size, popularity, and recency without globally sorting all objects.
\LHD{} (Figure~\ref{fig:lhd}) takes a different route.
It estimates each object's expected future \emph{return per cached byte-time} from online histograms and computes
\vspace{-1mm}
\[
  \text{hit density}
  = \frac{\text{hit probability}}{\text{object size} \times \text{expected time in cache}}
\vspace{-1mm}
\]
Thus, hit density can be read as expected future benefit per unit of cache byte-time, and objects with the smallest such return are evicted first.

\vspace{-2mm}
\subsection{Adaptive and Learned Policies}
These policies change their eviction signal online or learn a
predictor. They matter because cloud-analytics workloads often
alternate between scan-heavy, recency-dominated, and
frequency-dominated phases, so a fixed reuse model is often
insufficient.

\textit{\ARC{}, \LeCaR{}, and \Cacheus{}.}
\ARC{} (Figure~\ref{fig:arc}) keeps two resident lists, $T_1$ and $T_2$, and two ghost lists, $B_1$ and $B_2$.
$T_1$ stores objects seen only once recently, whereas $T_2$ stores objects seen at least twice recently.
$B_1$ and $B_2$ record recently evicted metadata from $T_1$ and $T_2$, respectively.
A hit in $B_1$ therefore indicates that a once-seen object was evicted too early and shifts capacity toward $T_1$, whereas a hit in $B_2$ indicates that a repeatedly reused object was evicted too early and shifts capacity toward $T_2$.
\LeCaR{} learns an online weight between two experts, \LRU{} and \LFU{}, using ghost-list feedback.
If a miss reveals that a previous eviction was a mistake, the expert responsible for that eviction is penalized.
\Cacheus{} keeps the same high-level meta-policy idea as \LeCaR{}, but replaces plain \LRU{} and \LFU{} with stronger experts: \emph{Scan-Resistant} \LRU{} (\textsf{SR-LRU}) and \emph{Churn-Resistant} \LFU{} (\textsf{CR-LFU}).
Its \textsf{SR-LRU} expert is specialized for scans and one-hit objects, and its \textsf{CR-LFU} expert is specialized for churn scenarios.
Operationally, each expert proposes a victim, and the meta-policy chooses between them according to the current online weights.

\textit{\LRB{}.}
\LRB{} (Figure~\ref{fig:lrb}) is a learned approximation to Belady-style eviction.
It trains a gradient boosting machine (GBM) to predict each object's time to next access from recent access-history features such as inter-reference distances, decaying counters, and object size.
At eviction time, it samples a subset of cached objects, predicts their next-use distances, and evicts the object predicted to be needed furthest in the future.

\textit{\GLCache{}.}
\GLCache{} reduces the cost of object-level prediction by learning at the group level instead.
It groups objects by insertion order, learns a utility score per group, and evicts from groups with lower expected benefit.
Compared with \LRB{}, the key idea is not finer prediction per object, but cheaper prediction shared across many similar objects.

\textit{\ThreeLCache{}.}
\ThreeLCache{} stays closer to \LRB{} because it still uses object-level learned prediction.
Its main goal is to reduce the overhead of that prediction while improving hit ratios through higher sampling quality.
First, it uses \emph{bidirectional sampling}, drawing candidates from both the tail of an \LRU{}-ordered queue, where stale objects are likely to sit, and the head, where newly admitted objects are likely to accumulate because \LRU{} places the most recent insertions near the head.
This makes the head a natural place to look for likely one-hit objects.
Second, it reuses prediction results across multiple evictions instead of inferring on every miss.
We evaluate two variants.
\ThreeLCache{}-BHR evicts by predicted next-use distance $D(o)$ alone, while \ThreeLCache{}-OHR uses:
\vspace{-1.5mm}
\[
  \text{Score}(o)=D(o)\times S(o)
\vspace{-1.5mm}
\]
This pair is useful because it isolates the effect of size normalization within the same learned framework.

\section{A Latency-Oriented Evaluation Metric}
\label{sec:metric-lens}

This section defines a latency-oriented evaluation metric for cache benefit in direct-access cloud OLAP.
A replacement policy directly controls the remote-download work caused by cache misses, not the full end-to-end query latency, yet directly measuring remote-download latency for every algorithm, workload, and cache size through live-cloud replay would be costly and noisy.
Thus, we need a simulation-friendly evaluation metric that remains faithful to remote-download latency while still separating the effects of request avoidance and byte avoidance.
As discussed in Section~\ref{subsec:metrics}, OHR captures request avoidance and BHR captures byte avoidance, but neither alone is sufficient in direct-access cloud OLAP because a miss to object storage combines a size-independent request overhead with a size-dependent transfer cost~\cite{durner2023exploiting}.
Delay-oriented metrics such as DSR~\cite{dsr} are closer in spirit, but they treat fetch delay as an empirical per-object quantity and do not expose this two-component miss structure.
This motivates a cloud-specific, latency-oriented extension of the hit-ratio family.

\vspace{-1mm}
\subsection{Cloud Download Latency Model}
\label{subsec:latency-model}

Our starting point is an explicit model of remote-download latency under direct-access cloud OLAP.
Following Durner et al.~\cite{durner2023exploiting}, we model the download latency of object $o$ with size $S(o)$ as:
\begin{equation}
\label{eq:cloud-latency}
L(o) = \alpha + \beta S(o)
\end{equation}
where $\alpha$ is the size-independent per-request latency and $\beta$ is the transfer latency per byte.
In our environment, fitting repeated S3 download measurements yields approximately $\alpha \approx 45$\,ms and $\beta \approx 12$\,ms/MiB.

This model makes the cloud miss-cost structure explicit.
When objects are small, the fixed term $\alpha$ dominates and reducing the number of misses matters most.
When objects are large, the transfer term $\beta S(o)$ dominates and reducing missed bytes becomes more important.
A latency-oriented cache metric for cloud OLAP should therefore reward both request avoidance and byte avoidance, with workload-dependent relative weights.
This model assumes independent requests and no bandwidth contention; we revisit its practical adequacy under live S3 replay in Section~\ref{subsec:uhr-validation}.

\vspace{-2mm}
\subsection{Unified Hit Ratio}
\label{subsec:uhr}

Let $\langle o_1,\ldots,o_{\Blue{N}}\rangle$ be the request sequence, let $S(o_i)$ denote the size of request $i$, and let $x_i \in \{0,1\}$ indicate whether request $i$ is served from cache.
We define:
\[
N_{\text{hit}} \equiv \sum_{i=1}^{\Blue{N}} x_i,\qquad
S_{\text{total}} \equiv \sum_{i=1}^{\Blue{N}} S(o_i),\qquad
S_{\text{hit}} \equiv \sum_{i=1}^{\Blue{N}} x_i S(o_i).
\]
Under the model above, a cache hit on request $i$ avoids remote-download latency $L(o_i)$. We therefore define \OURHR{} (\OURHRSHORT{}) as a latency-weighted hit ratio:
\begin{equation}
\label{eq:uhr-hit}
\OURHRSHORT{}
\equiv
\frac{\sum_{i=1}^{\Blue{N}} x_i L(o_i)}
     {\sum_{i=1}^{\Blue{N}} L(o_i)}
\end{equation}

This places \OURHRSHORT{} in the same family as OHR and BHR. OHR weights each hit by $1$, BHR weights each hit by $S(o_i)$, and \OURHRSHORT{} weights each hit by the remote-download latency it avoids, namely $L(o_i)=\alpha+\beta S(o_i)$. Substituting Eq.~\eqref{eq:cloud-latency} into Eq.~\eqref{eq:uhr-hit} yields:
\begin{equation}
\label{eq:uhr-expanded}
\OURHRSHORT{}
=
\frac{N_{\text{hit}}\alpha + S_{\text{hit}}\beta}
     {\Blue{N}\alpha + S_{\text{total}}\beta}
\end{equation}

Let $S_\mu \equiv S_{\text{total}}/\Blue{N}$ denote the mean request size.
The denominator of Eq.~\eqref{eq:uhr-expanded} factors as
$\Blue{N}(\alpha + \beta S_\mu)$, which makes the mean request size the natural scale for weighting OHR against BHR.  Then Eq.~\eqref{eq:uhr-expanded} can be rewritten as:
\begin{equation}
\label{eq:uhr-ohr-bhr}
\OURHRSHORT{}
=
w_O \cdot \mathrm{OHR} + w_B \cdot \mathrm{BHR}
\end{equation}
where
\begin{equation}
\label{eq:uhr-weights}
w_O = \frac{\alpha}{\alpha + \beta S_\mu},
\qquad
w_B = \frac{\beta S_\mu}{\alpha + \beta S_\mu}
\end{equation}

Eq.~\eqref{eq:uhr-ohr-bhr} makes the interpretation explicit: \OURHRSHORT{} is a workload-specific weighted combination of OHR and BHR, where the weights reflect the relative importance of request overhead and byte transfer. Equivalently, since cache hits are exactly the requests whose remote-download latency is avoided, \OURHRSHORT{} can also be viewed as the normalized fraction of aggregate remote-download latency saved by caching:
\begin{equation}
\label{eq:uhr-savings}
\OURHRSHORT{}
=
\frac{L_{\text{saved}}}{L_{\text{nocache}}}
=
1-\frac{L_{\text{remote}}}{L_{\text{nocache}}}
\end{equation}
where
\[
L_{\text{saved}} \equiv \sum_{i=1}^{\Blue{N}} x_i L(o_i),
\qquad
L_{\text{nocache}} \equiv \sum_{i=1}^{\Blue{N}} L(o_i)
\]

The weighted-hit interpretation in Eq.~\eqref{eq:uhr-hit} is our primary definition because it keeps \OURHRSHORT{} consistent with OHR and BHR, while the savings view in Eq.~\eqref{eq:uhr-savings} is useful when validating the metric against measured download latency. Unlike DSR~\cite{dsr}, \OURHRSHORT{} makes the request-versus-byte decomposition explicit, which is important for later analysis. Although Eq.~\eqref{eq:uhr-savings} shows that \OURHRSHORT{} is algebraically a normalized savings ratio, we retain the name \emph{Unified Hit Ratio} because Eq.~\eqref{eq:uhr-hit} defines it as a hit-ratio-style metric in which each hit is weighted by the latency it avoids.

\vspace{-2mm}
\subsection{Roles of OHR, BHR, and \OURHRSHORT{}}
\label{subsec:metric-roles}

We retain all three metrics because they answer complementary questions.
OHR measures request-count efficiency, including settings where many small misses or per-request object-store charges matter.
BHR measures byte efficiency when remote transfer volume dominates and \OURHRSHORT{} combines the two under the cloud miss-cost model and is therefore our primary proxy for remote-download latency.
\Blue{Accordingly, UHR is the direct metric for remote-download latency, OHR a metric for per-request charges such as S3 GET fees, and BHR a metric for per-byte transfer charges.}
Keeping all three is useful both for ranking and for explanation: a policy may rank highly on OHR because it is strongly size-aware, on BHR because it aggressively filters zero-reuse objects, or on \OURHRSHORT{} because it balances the two.
\Blue{We also note that \OURHRSHORT{} is designed to capture remote-download latency saved by caching rather than end-to-end query latency or local cache-access overhead.
In our measurements, local cache-access overhead is sub-ms and largely common across policies, whereas a 1\,MiB remote miss takes about 57\,ms.
Including this overhead would not materially change policy comparisons.}
Section~\ref{sec:empirical} first validates \OURHRSHORT{} against live S3 replay and then uses OHR, BHR, and \OURHRSHORT{} together to derive practical guidance.

\section{Empirical Study and Analysis}
\label{sec:empirical}

This section applies the evaluation metric of Section~\ref{sec:metric-lens} to a common direct-access setting and derives the evidence behind our practical guidance.
We proceed in four steps.
We first fix a common comparison setting so that all policies are evaluated under the same cache unit, miss-cost model, and eviction constraints.
We then perform a live-cloud reality check to verify that \OURHRSHORT{} tracks aggregate remote-download latency.
Next, we compare all policies exhaustively through trace-driven simulation and explain which workload traits favor which policy families.
Finally, we examine operation latency and metadata overhead.
The purpose of this section is not to announce a single universal winner, but to show which policy families work best under which workload characteristics and why.

\vspace{-1mm}
\subsection{Common Comparison Setting}
\label{subsec:common-setting}

To make the benchmark interpretable, we hold the cache unit, miss-cost model, and eviction constraints fixed across all policies.
We therefore evaluate every algorithm in the same direct-access physical-caching setting, using a common simulator, common traces, and common pin/unpin semantics.

\vspace{-1.5mm}
\subsubsection{Execution environment.}
All cloud-based experiments were run on a single Amazon EC2 m6i.metal instance with 128 CPU cores and 512\,GB of RAM, with data stored in Amazon S3 Standard.
To collect realistic object-reference traces, we executed workloads on SAP HANA (QRC 3/2025) configured to read data from S3.

\vspace{-1.5mm}
\subsubsection{\revision{Trace collection and simulation.}}
\label{sec:tracecollectionandsimulation}
\revision{During live trace collection, SAP HANA first identifies the data needed by each query.
It uses Parquet footer min/max statistics to prune irrelevant row groups and column projection to skip unreferenced columns before issuing cache requests.
We log the resulting physical column-chunk requests and pin/unpin events.
The simulator replays the same trace under every replacement policy.
All policies therefore see identical accesses and pinning constraints.}

\vspace{-1.5mm}
\subsubsection{Evaluated policies and implementation.}
\label{sec:libcachesim}
We evaluated all 22 policies in Table~\ref{tab:cache-algo-comparison} using libCacheSim~\cite{s3fifo,sieve,yang2023-qdlp}.
\Blue{The libCacheSim library serves as the common substrate for baseline policy implementations.
The HANA/S3 traces, simulator, UHR and its live-S3 validation, and the policy-selection analysis are specific to this study.
We extended the code with three algorithms absent from the original release, reimplemented one algorithm which had a correctness issue, and modified every policy's eviction logic to skip pinned objects.}
For learning-based baselines such as \LRB{}~\cite{lrb}, \GLCache{}~\cite{glcache}, and \ThreeLCache{}~\cite{3lcache}, we used the configurations recommended by the original authors.
\Blue{For \GDS{}, \GDSF{}, and \LFUDA{}, we evaluated both their standard uniform-cost form with $\mathit{Cost}(o)=1$ and a latency-cost variant with $\mathit{Cost}(o)=L(o)$ from Eq.~\eqref{eq:cloud-latency}.}
We also implemented \Crystal{}~\cite{durner2021crystal} as a semantic-caching baseline using the default RR/OR split and history settings from the original paper.
Counting the three latency-cost variants, the two \ThreeLCache{} variants, and the \Crystal{} baseline, the full evaluation covered 27 configurations in total, which is why the heatmaps in Figures~\ref{fig:ohr-heatmap}, \ref{fig:simulation_bhr}, and \ref{fig:simulation_uhr} each contain 27 columns.
Because our work assumes a direct-access physical-caching setting, we do not benchmark systems whose decisions are coupled to predicate pushdown, such as FlexPushdownDB~\cite{yang2021flexpushdowndb,yang2024flexpushdowndb} and LBSC~\cite{ji2024lbsc}.

\vspace{-1.5mm}
\subsubsection{Workloads.}
\label{sec:workloads}
\begin{figure*}[t]
  \centering
  \begin{minipage}[t]{\textwidth}
    \vspace{0pt}
    \centering
    \captionof{table}{Workload-specific UHR weights and \revision{column-chunk size variability. Median, CV, and p90/p10 are computed over distinct objects.}}
    \label{tab:uhr-weights}
    \vspace{-3mm}
    \small
    \setlength{\tabcolsep}{2pt}
    \begin{tabular}{lcccccc}
    \toprule
    Workload & {\makecell{$S_{\mu}$\\(MiB)}} & {$w_O$} & {$w_B$}
      & \revision{\makecell{Median\\(MiB)}} & \revision{CV} & \revision{p90/p10} \\
    \midrule
    CAB-MT    & 2.58 & 0.60 & 0.40 & \revision{1.51} & \revision{1.38} & \revision{18.85} \\
    CAB-MT2   & 2.58 & 0.60 & 0.40 & \revision{1.51} & \revision{1.37} & \revision{22.82} \\
    CAB-MT3   & 2.68 & 0.59 & 0.41 & \revision{1.51} & \revision{1.38} & \revision{18.85} \\
    CAB-ST    & 2.74 & 0.58 & 0.42 & \revision{1.51} & \revision{1.41} & \revision{18.85} \\
    MTD-DS    & 1.49 & 0.72 & 0.28 & \revision{1.28} & \revision{1.00} & \revision{37,250.89} \\
    Redbench  & 4.46 & 0.46 & 0.54 & \revision{0.77} & \revision{2.58} & \revision{1,941.71} \\
    Lineitem  & 1.54 & 0.71 & 0.29 & \revision{0.50} & \revision{1.20} & \revision{14.82} \\
    \bottomrule
    \end{tabular}
  \end{minipage}
  \par\vspace{3mm}
  \begin{minipage}[t]{0.8\textwidth}
    \vspace{0pt}
    \centering
    \setcounter{subfigure}{0}

    \includegraphics[width=0.9\linewidth]{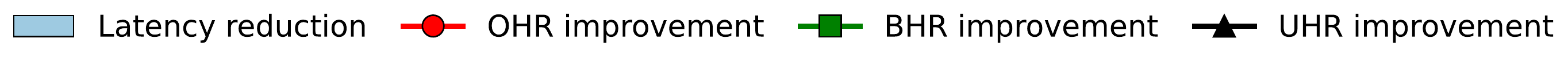}

    \begin{subfigure}[t]{0.49\linewidth}
      \centering
      \includegraphics[width=\linewidth]{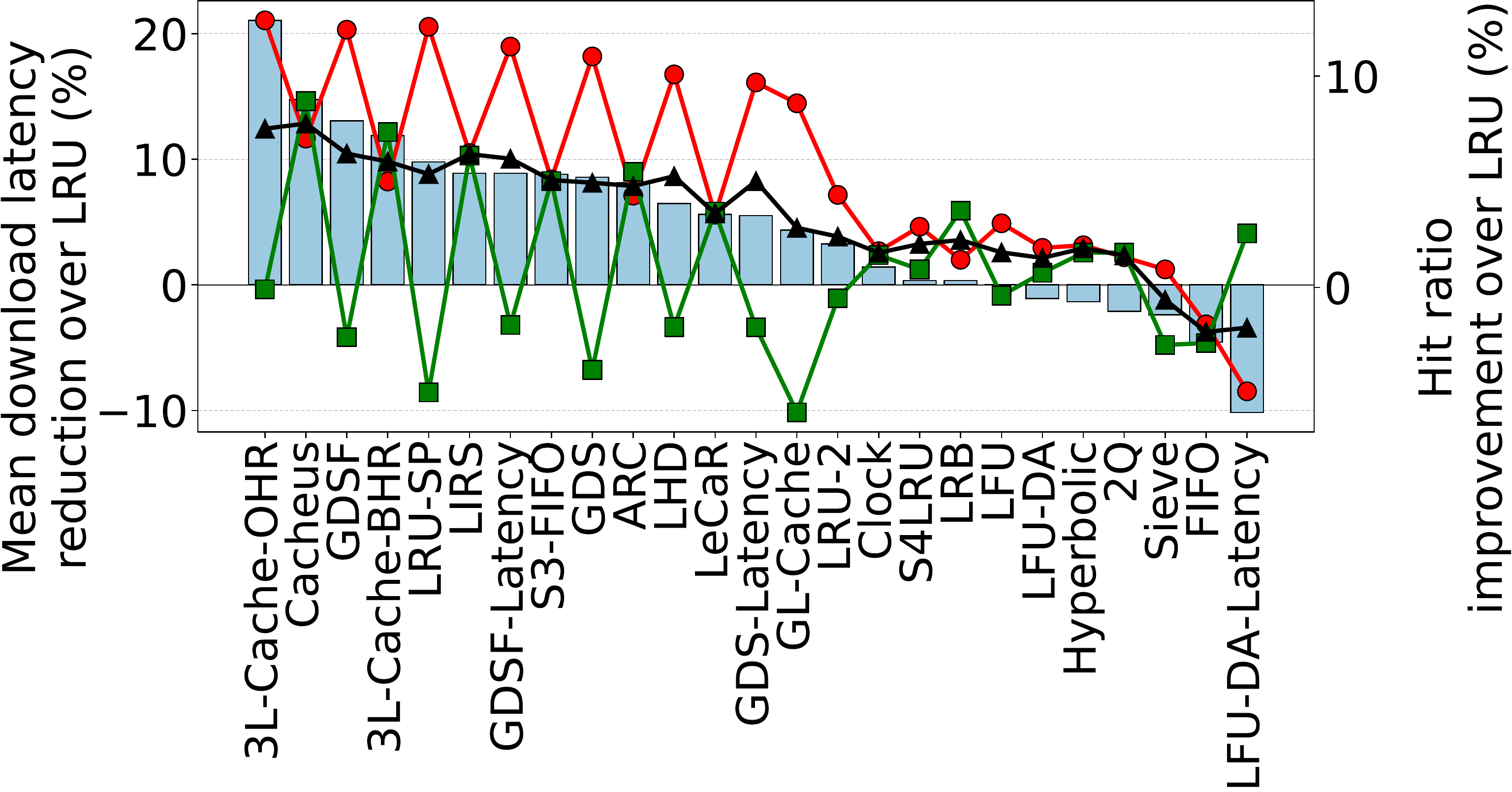}
      \vspace{-6mm}
      \caption{CAB-MT (20\% cache)}
    \end{subfigure}\hfill
    \begin{subfigure}[t]{0.49\linewidth}
      \centering
      \includegraphics[width=\linewidth]{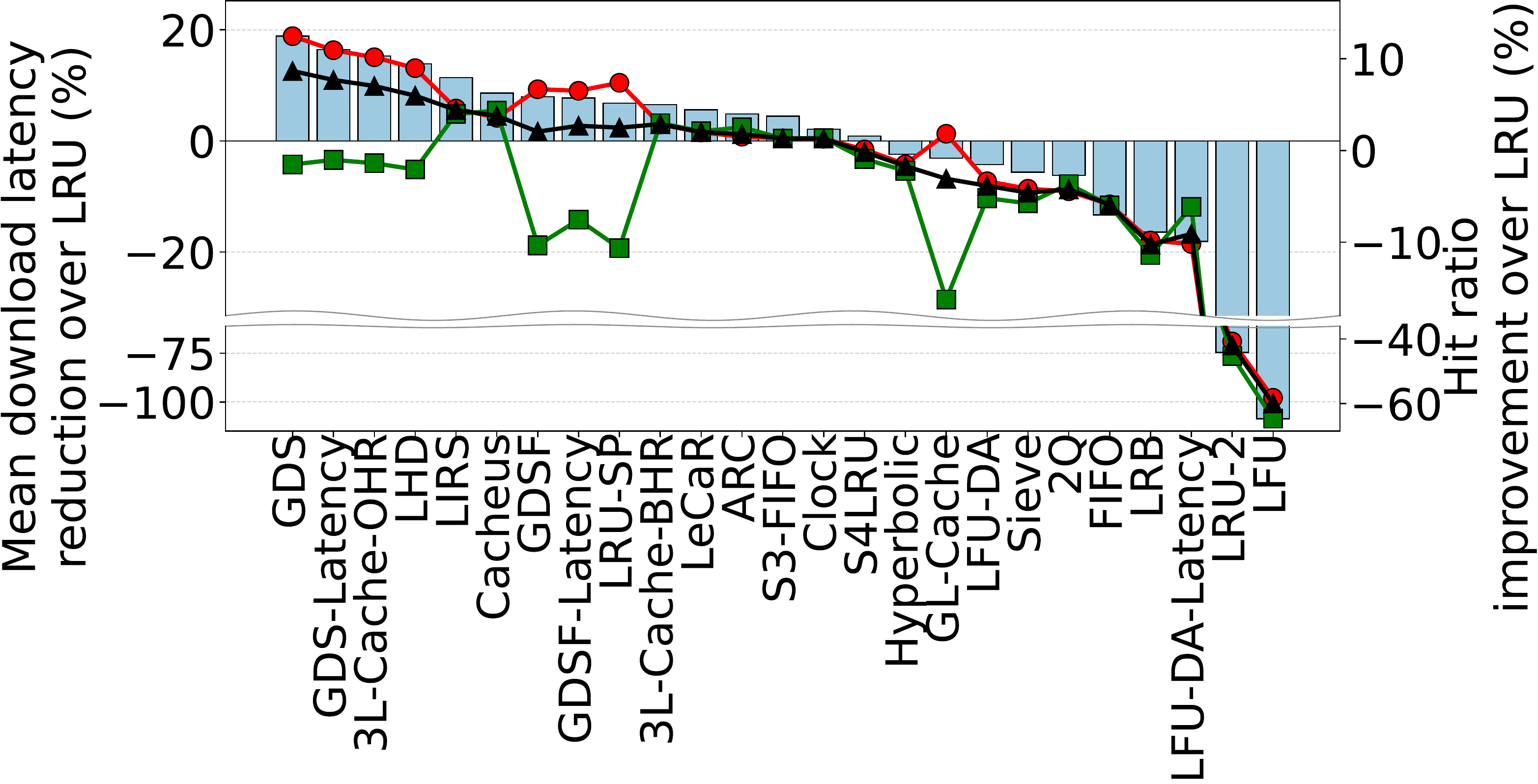}
      \vspace{-6mm}
      \caption{MTD-DS (5\% cache)}
    \end{subfigure}

    \vspace{-3mm}
    \captionof{figure}{Representative live-S3 validation. Improvements in \OURHRSHORT{}, OHR, and BHR (lines, right axis) versus reduction in aggregate remote-download latency over \LRU{} (bars, left axis), averaged over five runs.}
    \label{fig:uhr-validation}
  \end{minipage}
  \vspace{-8pt}
\end{figure*}

We use four workload families, yielding seven workload instances in total, to cover realistic analytical settings with both stable and shifting access patterns.
CAB~\cite{cab} is our main cloud-warehouse benchmark and provides four instances: CAB-MT, CAB-MT2, and CAB-MT3 are multi-tenant setups generated with different random seeds, while CAB-ST is a single-tenant setup.
Due to incomplete update support in the reference implementation~\cite{cab-github}, which leads to join inconsistencies, our evaluation focuses exclusively on read-only CAB workloads.
MTD-DS~\cite{yin2025mtd} extends TPC-DS to multi-tenant decision support and provides a broader query mix than CAB.
Redbench~\cite{redbench} synthesizes SQL workloads from anonymized cloud traces and preserves repetitions, bursty temporal structure, and mixed SELECT+DML behavior.
\Blue{Our Redbench instance contains approximately 36K statements, of which about 11K are DML operations. These operations consist primarily of small row-level updates, inserts, and deletes rather than bulk rewrites.
In our evaluated immutable-Parquet setting, SQL-level writes do not overwrite cached physical chunks in place. Instead, they create new physical chunks and make some previous chunks inactive for subsequent latest-snapshot reads. Across the Redbench trace, DML operations make 8,124 chunks (24.4 GiB) newly reachable and 8,126 chunks (24.8 GiB) inactive.}
Because \Crystal{} does not support table updates, we omit it from the Redbench evaluation and evaluate it on the remaining six read-only workloads.
Lineitem follows the selective region-reuse setting introduced by \Crystal{}~\cite{durner2021crystal} and serves as a favorable workload for semantic caching.
Table~\ref{tab:uhr-weights} summarizes the mean request size ($S_\mu$) and the resulting \OURHRSHORT{} weights ($w_O$, $w_B$) for all seven evaluated workloads.
\Blue{Counting each distinct object once, the median column-chunk size ranges from 0.50 to 1.51\,MiB, the coefficient of variation from 1.00 to 2.58, and the p90/p10 ratio from 14.82 to 37,250.
\Fluorescent{The largest p90/p10 ratios are driven by tiny lower-tail chunks, which make the p10 values very small.}
Column-chunk size variation reflects differences in column type, encoding, compression ratio, null distribution, and value distribution.
Because capacity is byte-limited while OHR counts every hit equally, size-aware policies can achieve higher OHR because retaining several small reusable chunks can yield more object hits than retaining a few large chunks.
}

\vspace{-1.5mm}
\subsubsection{Cache sizes}
\label{sec:cachesizes}
\Blue{Because workloads differ in working-set size and object-size distribution, the same absolute cache size would not impose comparable cache pressure for eviction across workloads.
Here, the cache pressure refers to the extent to which the cache is forced to evict objects because of the \Fluorescent{insufficient cache space} for the workload's active working set.
We therefore set a workload-specific cache-size range for fair comparisons of cache performance.
We determined the low end of this range as the point where the best policy reaches roughly 50\% \OURHRSHORT{}.
Similarly, we chose the high end as the point where it exceeded 90\% \OURHRSHORT{} and approached saturation.
}
We evaluated ten sizes in this range in simulation and three representative sizes in live-cloud validation.

\vspace{-1.5mm}
\subsubsection{Caching units}
\label{sec:cachingunits}
\Blue{Although the cache-size range is workload-specific, the cache unit is fixed as physical column chunks across all experiments.
All tables are stored in Apache Parquet using standard PyArrow default configurations, and the cache stores the physical column chunks requested by the query engine.
Fixing this unit across policies isolates replacement effects from cache-granularity effects.
Whole-file caching is a valid coarser alternative, but it would admit unprojected columns and pruned row groups, reducing byte efficiency and mixing granularity effects with replacement effects.}

Our trace-driven and single-flight live-replay methodology strictly follows the evaluation protocol established by foundational cache-replacement literature over the past two decades (e.g., \ARC{} \cite{arc}, \LRB{} \cite{lrb}, \SFIFO{} \cite{s3fifo}, \Sieve{} \cite{sieve}). Furthermore, our addition of live-S3 replay (Section~\ref{subsec:uhr-validation}) grounds this protocol in real-world latency dynamics, providing a more rigorous baseline than the simulation-only approaches typically used in these prior works.

\begin{table}[t]
\centering
\small
\setlength{\tabcolsep}{3pt}
\caption{Correlation between metric improvement and aggregate remote-download-latency reduction over LRU.
\Blue{For each workload/cache-size setting and metric, each policy configuration provides one data point $(x,y)$ where $x$ is the measured aggregate download-latency reduction over LRU, and $y$ is the corresponding metric improvement over LRU.
Pearson $r$ measures linear association, while Spearman $\rho$ measures rank agreement across policy configurations.}
Each cell reports ``$r$ / $\rho$''. Higher values indicate stronger association. Bold marks the strongest metric per row.}
\label{tab:latency-correlation}
\vspace{-3mm}
\begin{tabular}{llccc}
\toprule
\multirow{2}{*}{Workload} & \multirow{2}{*}{\makecell[l]{Cache\\Size}}
  & UHR & OHR & BHR \\
 & & ($r$ / $\rho$) & ($r$ / $\rho$) & ($r$ / $\rho$) \\
\midrule
\multirow{3}{*}{CAB-MT}
  & 10\% & \textbf{0.743 / 0.761} & 0.638 / 0.573 & 0.416 / 0.398 \\
  & 20\% & \textbf{0.946 / 0.968} & 0.834 / 0.865 & 0.149 / 0.133 \\
  & 30\% & \textbf{0.945 / 0.912} & 0.847 / 0.838 & 0.571 / 0.422 \\
\midrule
\multirow{3}{*}{MTD-DS}
  & 2\%  & \textbf{0.992 / 0.976} & 0.983 / 0.972 & 0.621 / 0.202 \\
  & 5\%  & \textbf{0.999 / 0.992} & 0.995 / 0.975 & 0.935 / 0.551 \\
  & 10\% & \textbf{0.999 / 0.972} & 0.995 / 0.882 & 0.989 / 0.832 \\
\midrule
\multirow{3}{*}{Lineitem}
  & 10\% & \textbf{1.000 / 0.914} & 0.992 / 0.913 & 0.800 / 0.803 \\
  & 15\% & \textbf{0.999 / 0.992} & 0.992 / 0.989 & 0.876 / 0.787 \\
  & 20\% & \textbf{0.998 / 0.989} & 0.978 / 0.962 & 0.837 / 0.767 \\
\bottomrule
\vspace{-10mm}
\end{tabular}
\end{table}

\vspace{-2mm}
\subsection{Reality Check: Validating \OURHRSHORT{} Against Live S3 Replay}
\label{subsec:uhr-validation}

Before using \OURHRSHORT{} in the broad comparison, we verify that it tracks the quantity it is meant to approximate: aggregate remote-download latency.
Here, aggregate remote-download latency means the sum of S3 fetch latencies over all cache misses during a replay, and we report the mean of this aggregate quantity over repeated runs.
This validation is intentionally limited to three workload families and three cache-size points, because its purpose is metric validation rather than exhaustive policy ranking.
\Blue{The exhaustive comparison across all seven workload instances is instead performed by trace-driven simulation in Section~\ref{subsec:broad-comparison}.}
We use CAB-MT, MTD-DS, and Lineitem because together they span workloads of mixed request/byte levels, a more request-dominated state, and a highly selective region-reuse pattern.

Using the request traces collected from CAB-MT, MTD-DS, and Lineitem, we replay requests at their original arrival rates through our cache framework while serving misses by actual downloads from Amazon S3.
We evaluate three representative cache sizes per workload, and repeat each replay five times to reduce cloud-network noise.
While the model in Section~\ref{subsec:latency-model} assumes independent requests and no bandwidth contention, these live replays expose policies to real S3 timing variability and therefore serve as a practical reality check.
\Blue{For each configuration, we compare its OHR, BHR, and UHR improvements over LRU with the corresponding reduction in aggregate remote-download latency.}
Figure~\ref{fig:uhr-validation} shows representative CAB-MT and MTD-DS cases, while Table~\ref{tab:latency-correlation} summarizes all nine validated configurations.
\Blue{In Figure~\ref{fig:uhr-validation}, policies are sorted in decreasing order of the measured latency reduction.
A metric is considered well-aligned with the measured reduction when its line decreases monotonically along this order.}

\begin{figure}[t]
  \centering
  \includegraphics[width=0.6\linewidth]{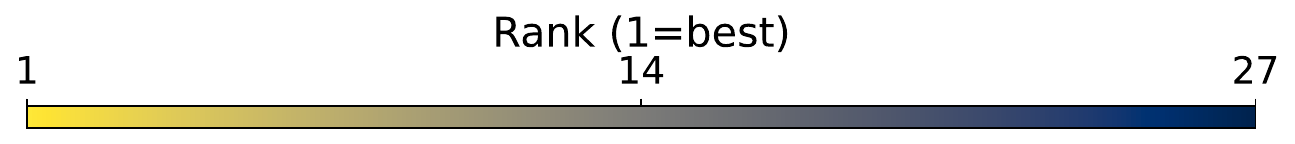}
  \includegraphics[width=0.75\linewidth]{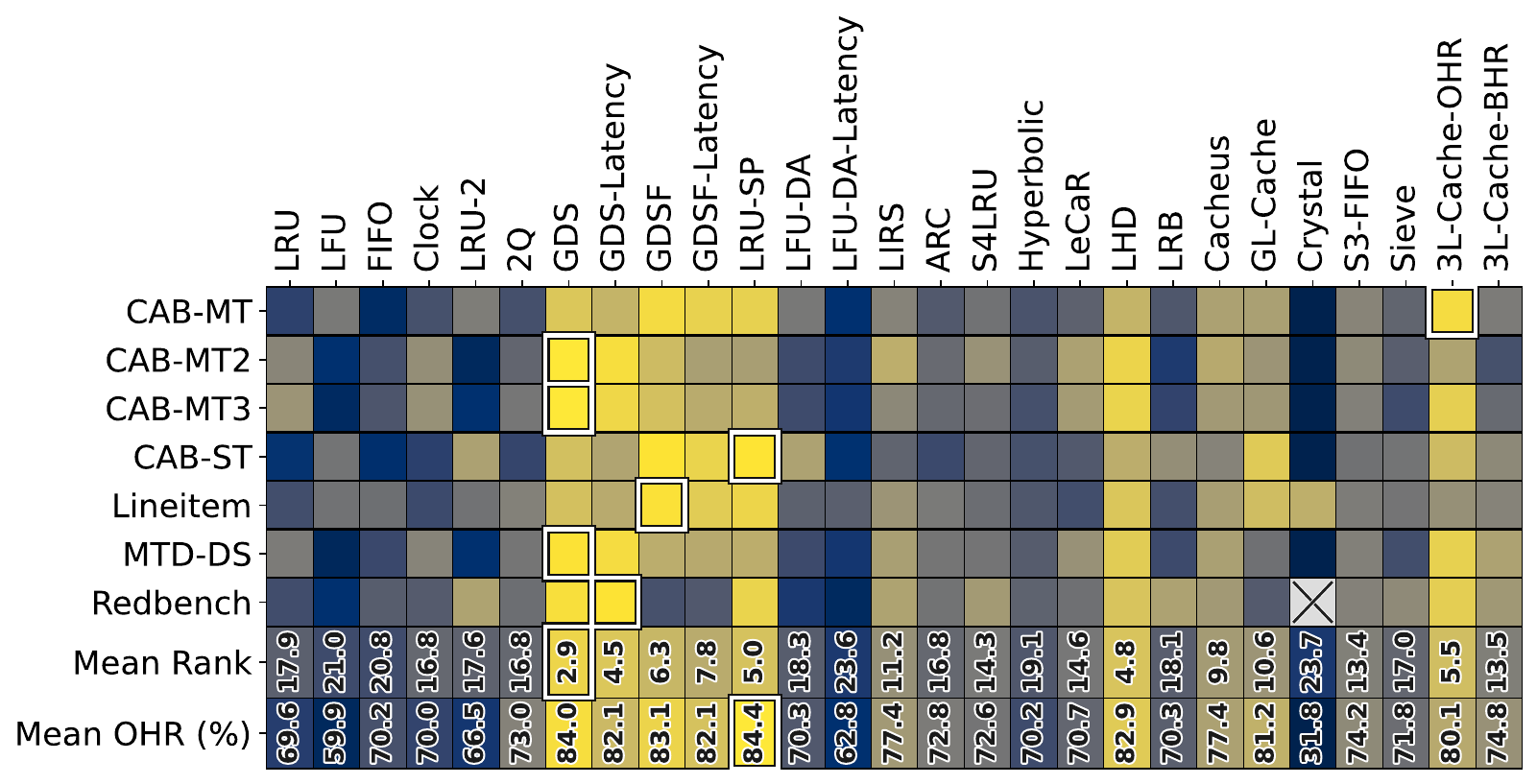}

  \vspace{-9.5pt}
  \caption{\revision{Mean OHR rank heatmap.
  Within each workload, policies are ranked separately at each of ten cache sizes, and the ten ranks are averaged. The brighter cell color indicates \Fluorescent{a better/lower rank}.}
  Bottom two rows report the overall mean rank and mean OHR.}
  \vspace{-18pt}
  \label{fig:ohr-heatmap}
\end{figure}

The validation results support \OURHRSHORT{} as the primary latency-oriented objective.
\Blue{
Table~\ref{tab:latency-correlation} shows that UHR has the strongest Pearson and Spearman correlations with measured latency reduction across policy configurations, across all nine validation settings, although its margin over OHR is small in MTD-DS.
This small margin is expected because the fixed request term gives OHR a large weight in UHR for MTD-DS.}
\Blue{In CAB-MT, the Pearson and Spearman correlations between UHR improvement and measured remote-download-latency reduction are lower overall than in MTD-DS and Lineitem.
This is because UHR varies less across the evaluated policies on CAB-MT.
The standard deviation of UHR across policies is 0.038, 0.025, and 0.016 at the three CAB-MT cache sizes, respectively, compared with 0.076--0.107 for MTD-DS and 0.230--0.293 for Lineitem.
Because UHR differs little across policies on CAB-MT, fluctuations in S3 fetch latency during live replay can obscure the small policy-level differences and lower the observed correlations.
UHR nevertheless remains the strongest metric in every CAB-MT row.}

\Blue{
We additionally assess latency-model sensitivity using 220 one-hour CAB-MT live-replay runs collected over 10 days.
Across these runs, the fitted $\alpha$ ranges from 33.5 to 65.1\,ms and $\beta$ from 10.1 to 15.7\,ms/MiB.
OHR and BHR do not depend on these parameters.
For UHR, $\alpha$ and $\beta$ enter only through the weights $w_O$ and $w_B$ in Eq.~\eqref{eq:uhr-weights}.
For a given mean request size $S_\mu$, these weights depend only on the ratio $\alpha/\beta$ rather than on the individual values of $\alpha$ and $\beta$.
A larger ratio shifts UHR toward OHR, while a smaller ratio shifts it toward BHR.
In our runs, the fitted $\alpha$ and $\beta$ co-vary (Pearson $r=0.62$), which keeps the ratio stable.
As a result, $w_O$ stays within 0.54--0.62, compared with 0.60 under the reported parameters.
Recomputing UHR with the run-specific fits changes it by at most 0.005, with a mean change of 0.0007.
}

The key point is not merely that \OURHRSHORT{} correlates with latency, but that it remains interpretable: later sections can attribute high \OURHRSHORT{} either to strong request avoidance, to strong byte avoidance, or to a workload-dependent balance of both.
We therefore use \OURHRSHORT{} as our primary proxy for remote-download latency in the large-scale comparison, while retaining OHR and BHR for explanation.

\vspace{-2mm}
\subsection{Broad Comparison Across Workloads and Cache Sizes}
\label{subsec:broad-comparison}

\begin{figure}[t]
  \centering
    \includegraphics[width=0.6\linewidth]{experiment_results/simulation_hit_ratio_results/heatmap_plots_civids/legend_no_lara.pdf}
    {\includegraphics[width=0.75\linewidth]{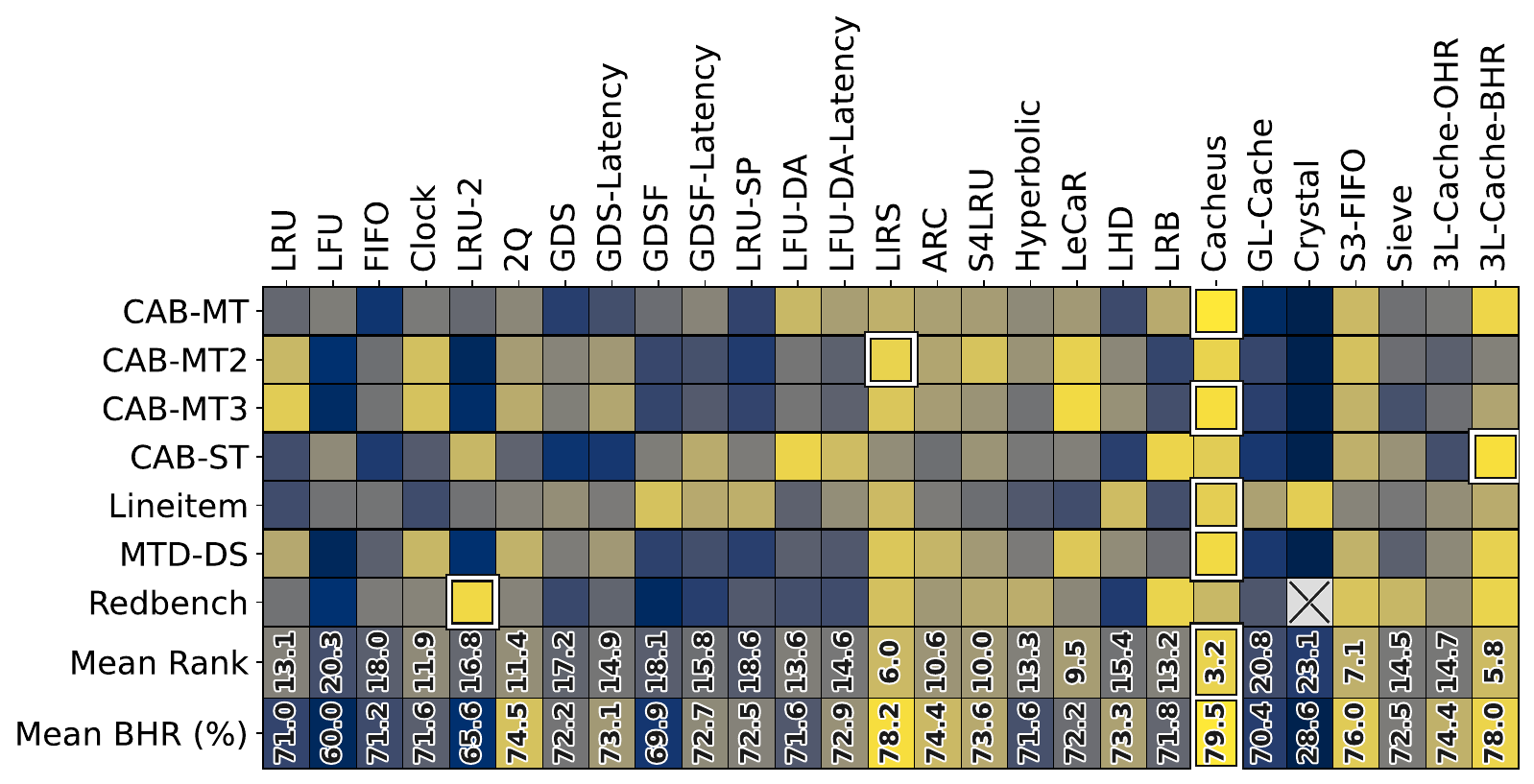}
    }
  \vspace{-4.5pt}
  \caption{\revision{Mean BHR rank heatmap.
  Within each workload, policies are ranked separately at each of ten cache sizes, and the ten ranks are averaged. The brighter cell color indicates \Fluorescent{a better/lower rank}.}
  Bottom two rows report the overall mean rank and mean BHR.}
  \vspace{-13pt}
  \label{fig:simulation_bhr}
\end{figure}

\begin{figure}[t]
  \centering
  \includegraphics[width=0.7\linewidth]{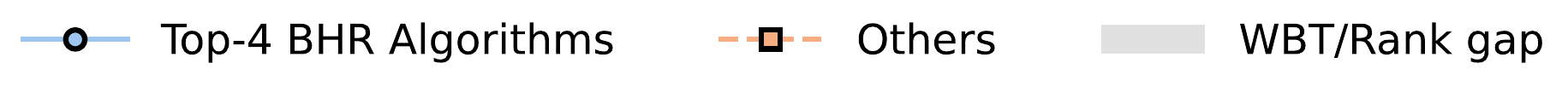}
  
  \begin{subfigure}[t]{0.49\columnwidth}
    \centering
    \includegraphics[width=\linewidth]{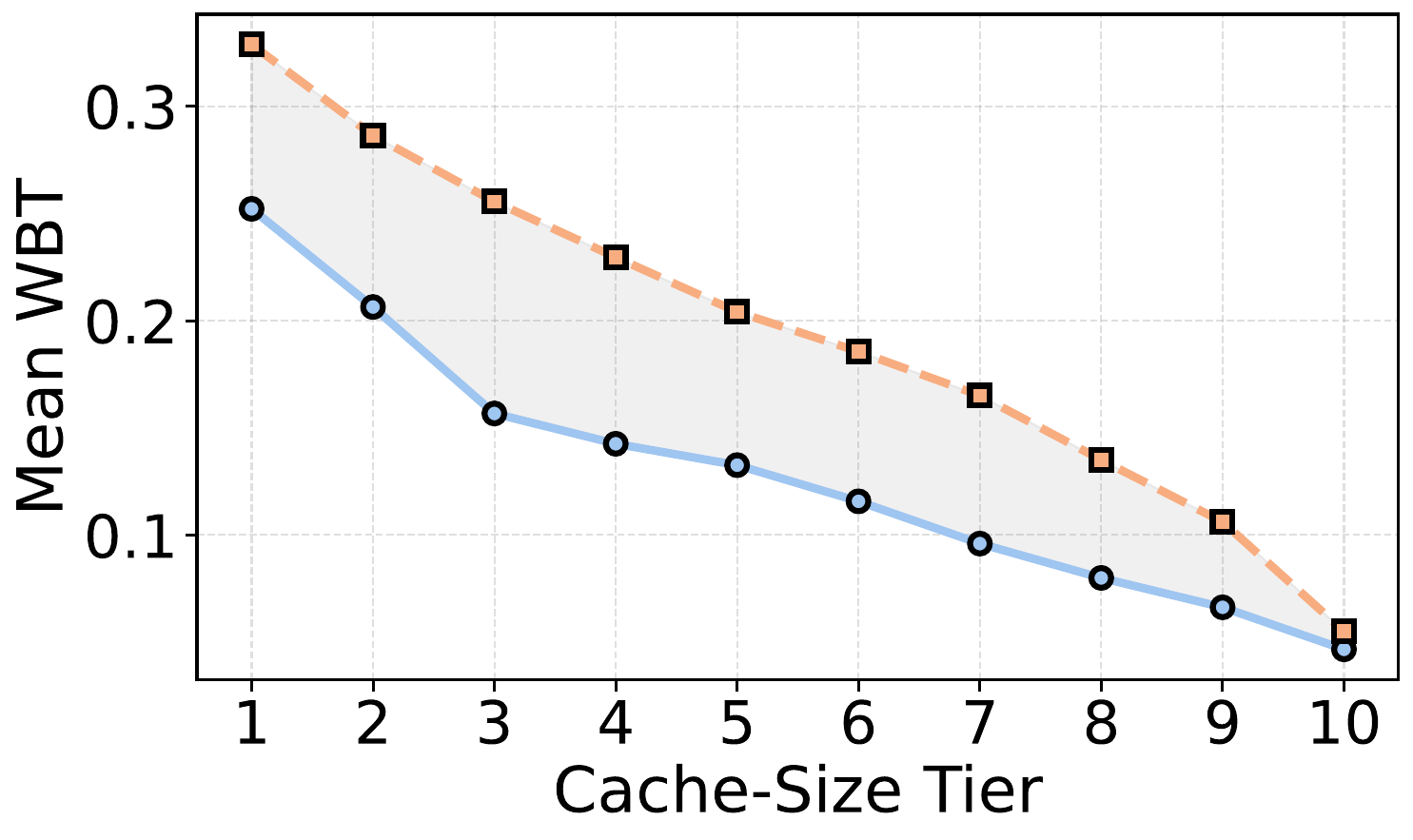}
    \vspace{-6mm}
    \caption{\footnotesize Mean WBT by cache-size tier.}
    \label{fig:wbt-tier}
  \end{subfigure}\hfill
  \begin{subfigure}[t]{0.49\columnwidth}
    \centering
    \includegraphics[width=\linewidth]{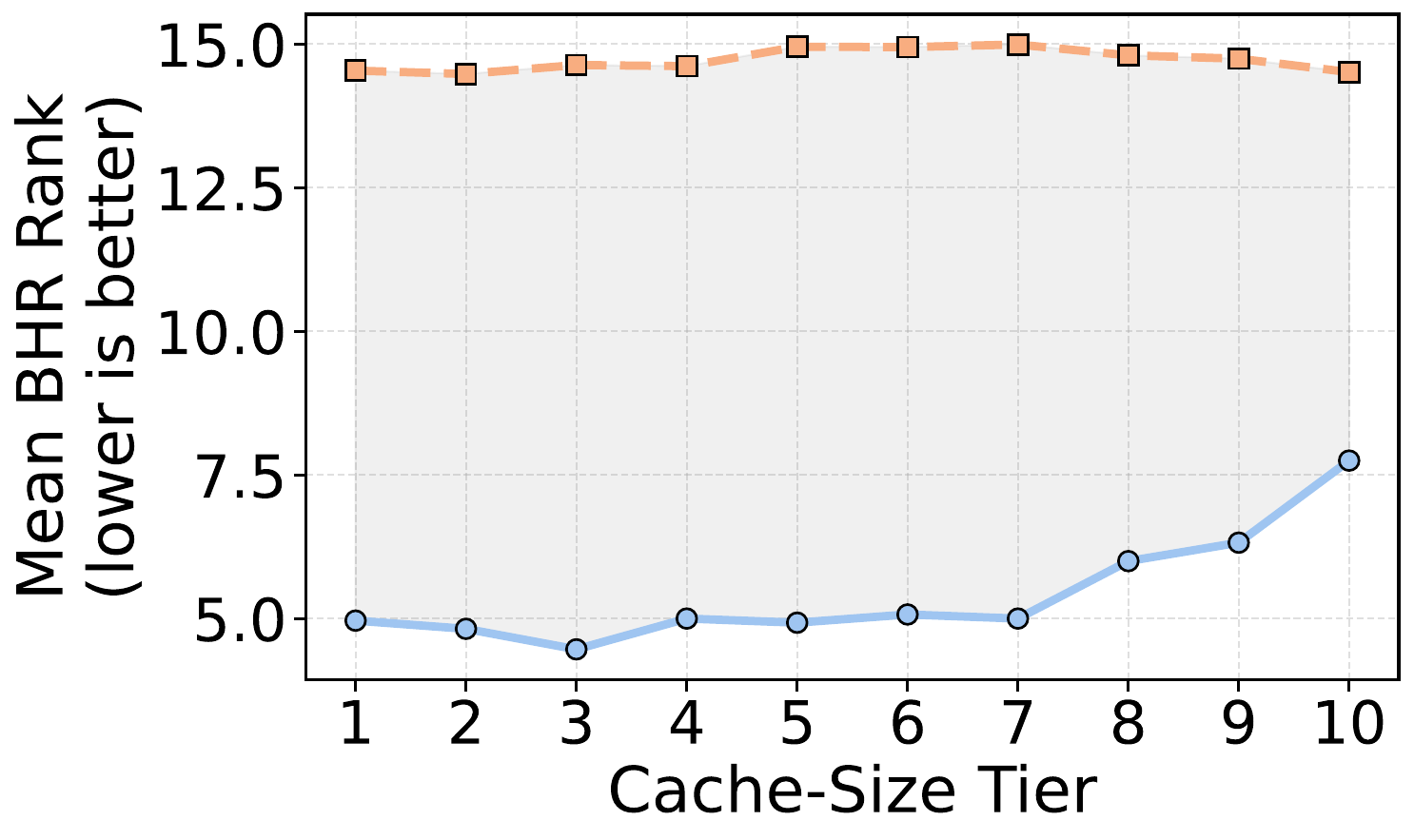}
    \vspace{-6mm}
    \caption{\footnotesize Mean BHR rank by cache-size tier.}
    \label{fig:bhr_top4_rank_over_cache_size}
  \end{subfigure}
  
  \vspace{-4mm}
  \caption{WBT and BHR rank by cache-size tier. Each workload's ten evaluated cache sizes (Section~\ref{subsec:common-setting}) are numbered from tier~1 (smallest) to tier~10 (largest).}
  \label{fig:wbt_analysis}
  \vspace{-4mm}
\end{figure}

\begin{figure}[t]
  \centering
    \includegraphics[width=0.6\linewidth]{experiment_results/simulation_hit_ratio_results/heatmap_plots_civids/legend_no_lara.pdf}
    {\includegraphics[width=0.75\linewidth]{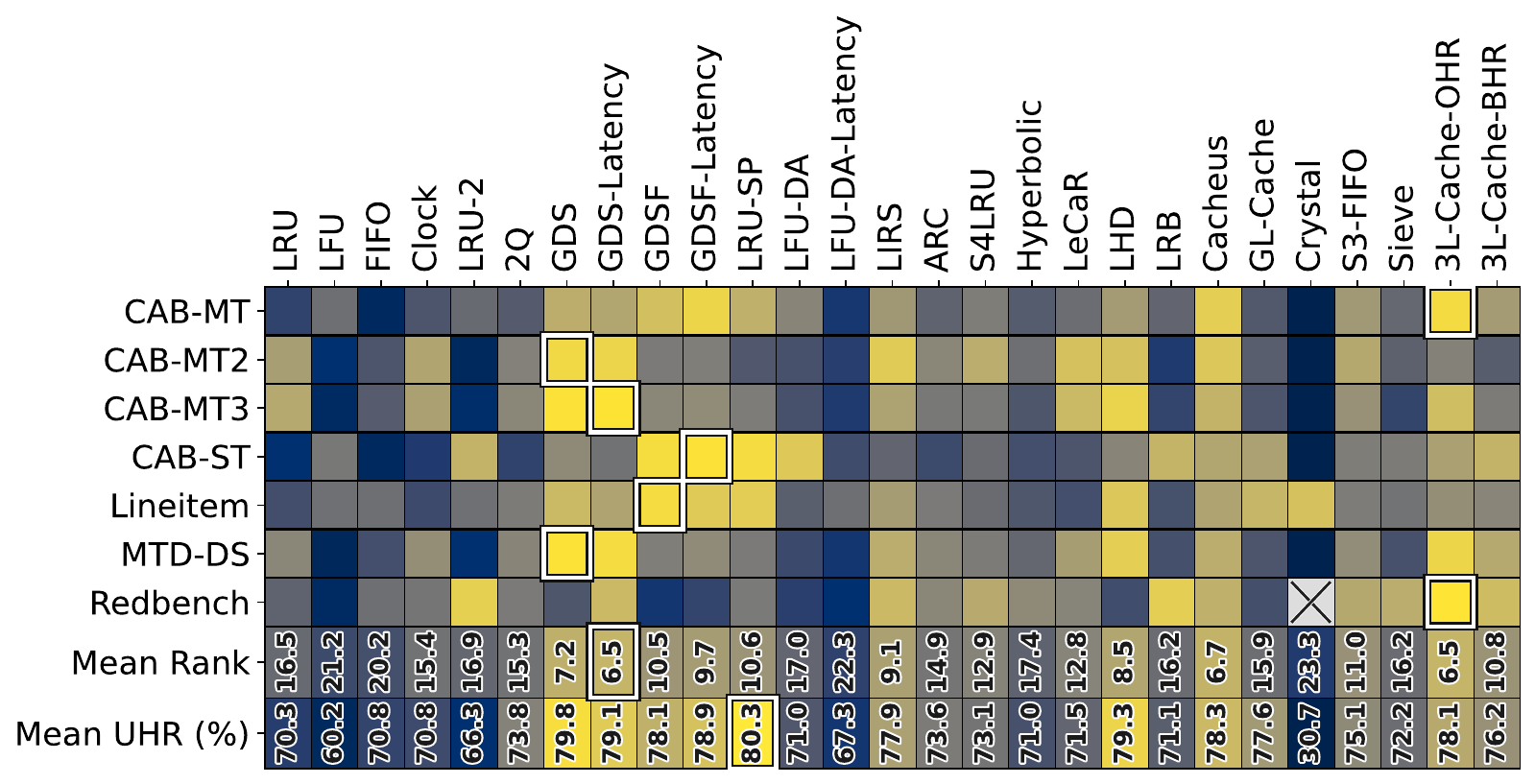}
    }
  \vspace{-7pt}
  \caption{\revision{Mean \OURHRSHORT{} rank heatmap.
  Within each workload, policies are ranked separately at each of ten cache sizes, and the ten ranks are averaged. The brighter cell color indicates \Fluorescent{a better/lower rank}.}
  Bottom two rows report the overall mean rank and mean \OURHRSHORT{}.}
  \vspace{-14pt}
  \label{fig:simulation_uhr}
\end{figure}

\begin{figure}[t]
    \centering
    \includegraphics[width=0.7\linewidth]{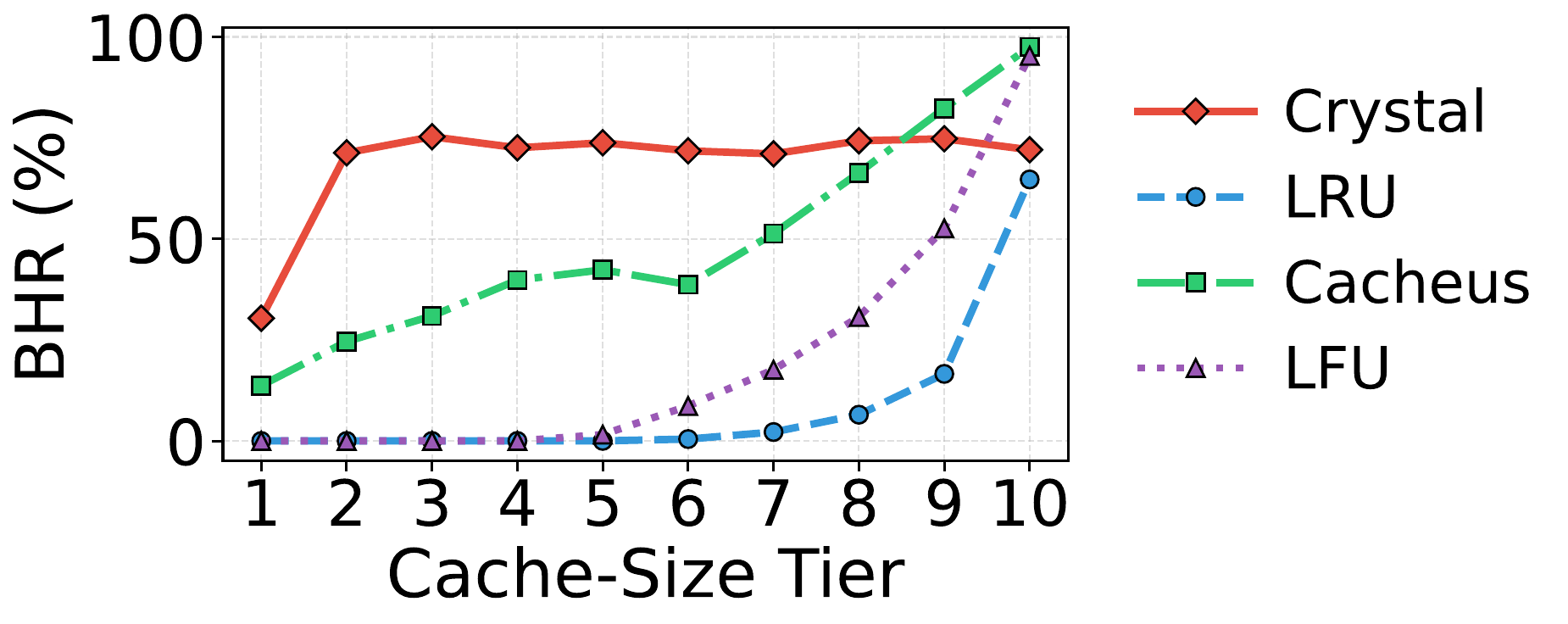}
    \vspace{-13pt}
    \caption{BHR of Crystal, Cacheus, \LRU{}, and \LFU{} on the Lineitem workload across all cache-size tiers. Ten evaluated cache sizes (Section~\ref{subsec:common-setting}) are numbered from tier~1 (smallest) to tier~10 (largest).}
    \vspace{-13pt}
    \label{fig:crystal-lineitem}
\end{figure}

With the evaluation metric validated, we now compare all 27 configurations across the full workload and cache-size space.
We evaluate the three metrics OHR, BHR, and \OURHRSHORT{} through trace-driven simulation to achieve exhaustive coverage and consistency across experiments.
Figures~\ref{fig:ohr-heatmap}, \ref{fig:simulation_bhr}, and \ref{fig:simulation_uhr} present per-workload rank heatmaps (1 = best) for each metric, aggregated over ten cache sizes per workload.
\Blue{For each metric, we rank all policies independently at each workload/cache-size pair and then average the ten cache-size-specific ranks for each workload-policy pair.
This rank-based aggregation avoids averaging raw hit ratios whose absolute values change with cache pressure.
Since policy rankings can vary across cache sizes, we report key cache-size-specific results separately in Figures~\ref{fig:wbt_analysis} and~\ref{fig:crystal-lineitem}.
}
Therefore, we do not treat aggregate mean ranks as the sole evidence for any design claim, which is especially important because several workload instances belong to the same CAB family.
Overall, the strongest OHR and BHR policies rarely coincide.
Size-awareness generally gives strong OHR, whereas effective zero-reuse filtering is needed for strong BHR.

\textit{Impact of write-driven churn.}
\Blue{As described in Section~\ref{subsec:common-setting}, Redbench DML operations change the active physical-chunk set and therefore act as workload drift rather than as separate invalidation events.
OHR, BHR, and UHR reflect the resulting compulsory misses and loss of future hit opportunities, while the qualitative policy-family trends on Redbench remain consistent with the broader comparison.}

\textit{When size-aware policies win.}
Size-aware policies dominate OHR because, under a fixed cache budget, they retain more objects and therefore create more opportunities for request hits.
This pattern is clear in Figure~\ref{fig:ohr-heatmap}: \GDS{}, \GDSF{}, \LRUSP{}, \LHD{}, and \ThreeLCache{}-OHR consistently occupy the strongest positions.
Pairwise comparisons isolate the mechanism.
\GDSF{} differs from \LFUDA{} mainly by dividing $\mathit{Freq}(o)\cdot \mathit{Cost}(o)$ by object size, and \ThreeLCache{}-OHR differs from \ThreeLCache{}-BHR mainly by adding size normalization to the same learned framework.
In both cases, the size-normalized variant ranks substantially better on OHR.
A notable result is that classical size-aware policies such as \GDS{} and \GDSF{} often outperform recent learned baselines.

\emph{$\square$ Practical takeaway: When the main objective is request avoidance under variable-sized objects, size normalization is the dominant design choice.}

\begin{figure}[t]
  \centering

  \begin{minipage}[c]{0.7\linewidth}
    \centering
    \includegraphics[width=\linewidth]
      {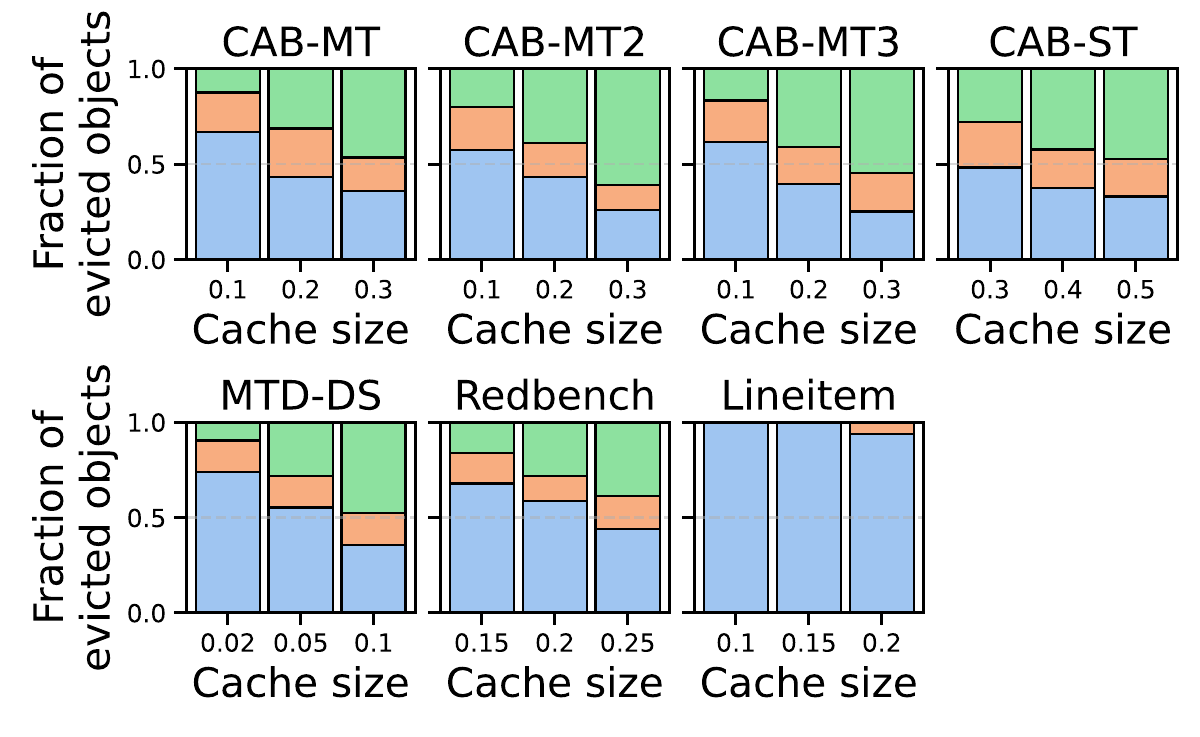}
  \end{minipage}\hfill
  \begin{minipage}[c]{0.2\linewidth}
    \centering
    \includegraphics[width=\linewidth]
      {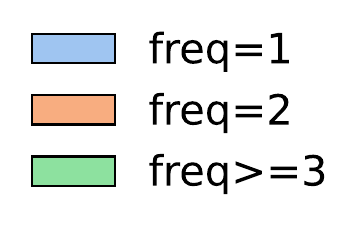}
  \end{minipage}

  \vspace{-6mm}
  \caption{Frequency of evicted objects under \LRU{} across workloads and
  cache sizes. $\text{Freq}=1$ objects received no re-access after admission.}
  \label{fig:lru-evict-freq}
  \vspace{-15pt}
\end{figure}

\textit{When zero-reuse filtering matters.}
BHR is driven less by size normalization than by early removal of objects that are admitted once and never hit again.
We call an object \emph{zero-reuse} if it receives no cache hit between admission and eviction.
Because a hit on an object of size $S(o)$ contributes $S(o)$ bytes to BHR but the same object also occupies $S(o)$ bytes of cache space, size normalization does not automatically improve byte efficiency.
Instead, as shown in Figure~\ref{fig:simulation_bhr}, the strongest BHR policies---notably \Cacheus{}, \SFIFO{}, \LIRS{}, and \ThreeLCache{}-BHR---share aggressive zero-reuse filtering, although they realize it through different mechanisms.

To quantify the waste caused by zero-reuse objects, we define the wasted byte-time fraction (WBT):
\begin{equation}
\label{eq:wbt}
\mathrm{WBT}
=
\frac{\sum_{j \in E_0} S(o_j)\cdot \tau_j}{C\cdot T}
\end{equation}
where $E_0$ is the set of zero-reuse evictions, $\tau_j$ is the residency time of object $o_j$, $C$ is cache capacity, and $T$ is trace length.
Intuitively, WBT is the ratio of byte-time consumed by zero-reuse objects to the total cache byte-time budget, since each such object occupies cache resources in proportion to both its size and its residency time.
The denominator $C\cdot T$ is the total \emph{cache byte-time budget}: the amount of byte-capacity available over the full trace.
Figure~\ref{fig:lru-evict-freq} shows that under \LRU{}, a large fraction of evicted objects are one-hit across all workloads.
Figures~\ref{fig:wbt-tier} and~\ref{fig:bhr_top4_rank_over_cache_size} show that the separation between the top BHR policies and the rest is strongest at small cache sizes.

However, low WBT is necessary but not sufficient for strong BHR.
\LFU{} provides the counterexample: it can achieve relatively low WBT by deprioritizing frequency-1 objects, yet still ranks poorly on BHR because stale frequency counts degrade allocation quality among reusable objects.
At larger cache sizes, WBT converges across algorithms but BHR gaps persist, indicating that allocation quality among reusable objects becomes the main differentiator.

\emph{$\square$ Practical takeaway. BHR depends critically on aggressive zero-reuse filtering under high cache pressure, and on good allocation quality among reusable objects once that pressure relaxes.}

\textit{When latency-cost scoring helps---and when it does not.}
\OURHRSHORT{} rewards policies that balance request avoidance and byte avoidance under workload-specific weights.
Adding a \Blue{latency-cost} term helps only when the underlying priority formulation already handles object size correctly.
As shown in Figure~\ref{fig:simulation_uhr}, \GDS{}-Latency and \GDSF{}-Latency improve over their uniform-cost counterparts, whereas \LFUDA{}-Latency worsens because it multiplies by latency without size normalization and therefore overprotects large objects.
In mean rank, GDS-Latency ties with \ThreeLCache{}-OHR for the strongest \OURHRSHORT{} result at 6.5, followed closely by \Cacheus{} at 6.7.
More broadly, the strongest \OURHRSHORT{} policies do not form a separate family.
The \OURHRSHORT{} frontier contains both OHR-strong size-aware policies such as \LRUSP{} and \ThreeLCache{}-OHR and BHR-strong filtering-oriented policies such as \Cacheus{} and \LIRS{}.

\emph{$\square$ Practical takeaway: Under \OURHRSHORT{}, policies that are already strong on either OHR or BHR meet on a common frontier, and \Blue{latency-cost} injection helps only when the underlying formula preserves a good size-versus-reuse trade-off.}

\begin{figure}[t]
  \centering
    \includegraphics[width=0.6\linewidth]{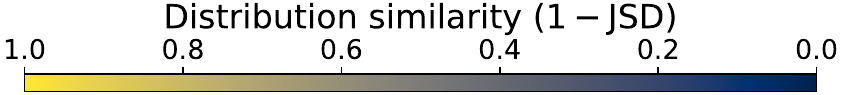}
    {\includegraphics[width=0.75\linewidth]{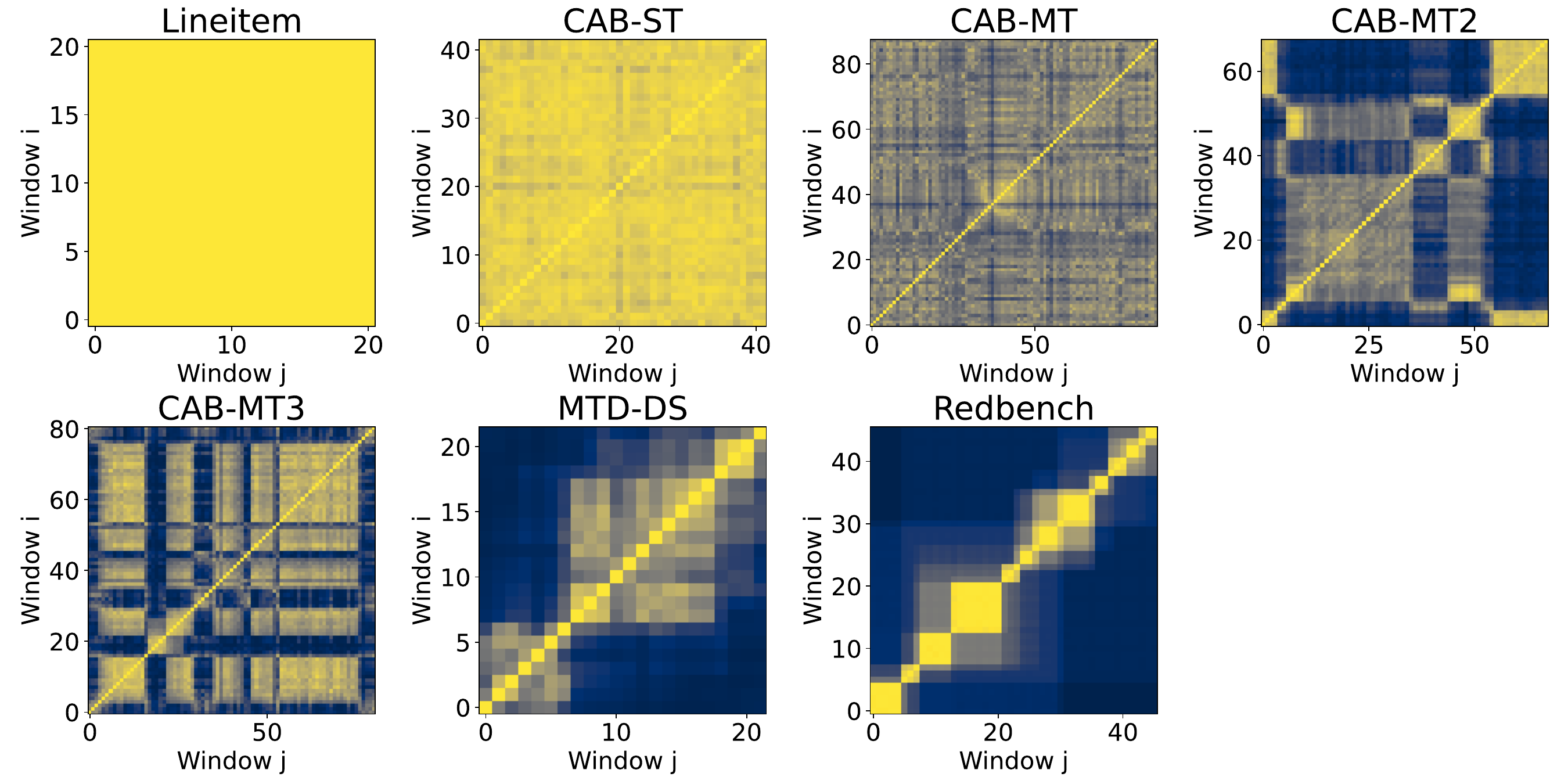}
    }
  \vspace{-9pt}
  \caption{Pairwise Jensen-Shannon similarity between fixed-size request windows ($W = 32$K requests per window) for each workload.
  Cell $(i, j)$ reports $1 - \mathrm{JSD}(p_i, p_j)$, where $p_i$ is the object-level request-share distribution in window $i$.
   Bright cells indicate similar distributions between two windows; dark cells indicate distributional shift. Other window sizes such as $W = 8$K and $W = 128$K yield similar patterns.}
  \vspace{-16pt}
  \label{fig:jsd_similarity}
\end{figure}

\textit{When online adaptivity pays off.}
Adaptivity matters because the value of recency and frequency changes across workloads and over time.
Stationary workloads favor accumulated frequency, whereas non-stationary workloads favor recency.
The same contrast appears not only between \LFU{} and \LRU{}, but also between \GDSF{} and \GDS{}: adding frequency helps when reuse is stable and hurts when frequency history becomes stale.
Figure~\ref{fig:jsd_similarity} shows the pairwise Jensen-Shannon similarity between fixed-size request windows of $W = 32$K requests, where bright cells indicate similar request distributions and dark cells indicate dissimilar ones.
Lineitem is perfectly stationary and CAB-ST near-stationary, while the remaining five workloads exhibit clear distribution shifts.

Adaptive policies improve robustness by changing their behavior online rather than committing to a fixed recency--frequency balance.
\ARC{} and \LeCaR{} already improve over static baselines, but \Cacheus{} is stronger because it combines two specialized experts: one that reacts quickly to scans and phase shifts, and one that preserves stronger reuse history.
Learned policies can also adapt in principle, but their retraining cadence and history windows are coarser than per-request meta-algorithm feedback.
Under rapidly shifting multi-tenant workloads, this lag limits their advantage.

\emph{$\square$ Practical takeaway: When workloads drift, prefer lightweight online adaptation over a fixed recency--frequency balance, and use retraining-heavy learned methods selectively.}

\textit{When semantic caching helps.}
Semantic caching is strongest when queries repeatedly target highly selective, region-like predicates whose qualifying tuples are sparse in the underlying table.
In this setting, as illustrated by the Lineitem workload, \Crystal{} can achieve high BHR at very small cache sizes because it materializes only qualifying tuples rather than full column chunks (Figure~\ref{fig:crystal-lineitem}).
Outside this access pattern, \Crystal{} often underperforms well-tuned chunk caching.
Its materialized regions are coarse when later queries request partially overlapping columns, and its OR-cache refresh operates in bulk and therefore adapts slowly to workload shifts.

\emph{$\square$ Practical takeaway: Semantic caching is a specialized option for selective region-like reuse, not a general replacement for well-tuned column-chunk caching.}

\begin{figure}[t]
    \centering
    \includegraphics[width=\linewidth]{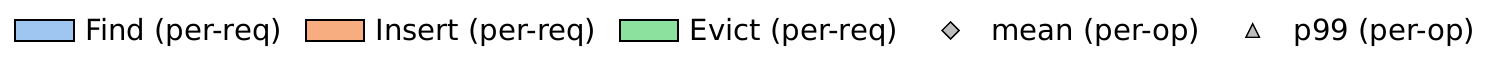}
    \includegraphics[width=\linewidth]{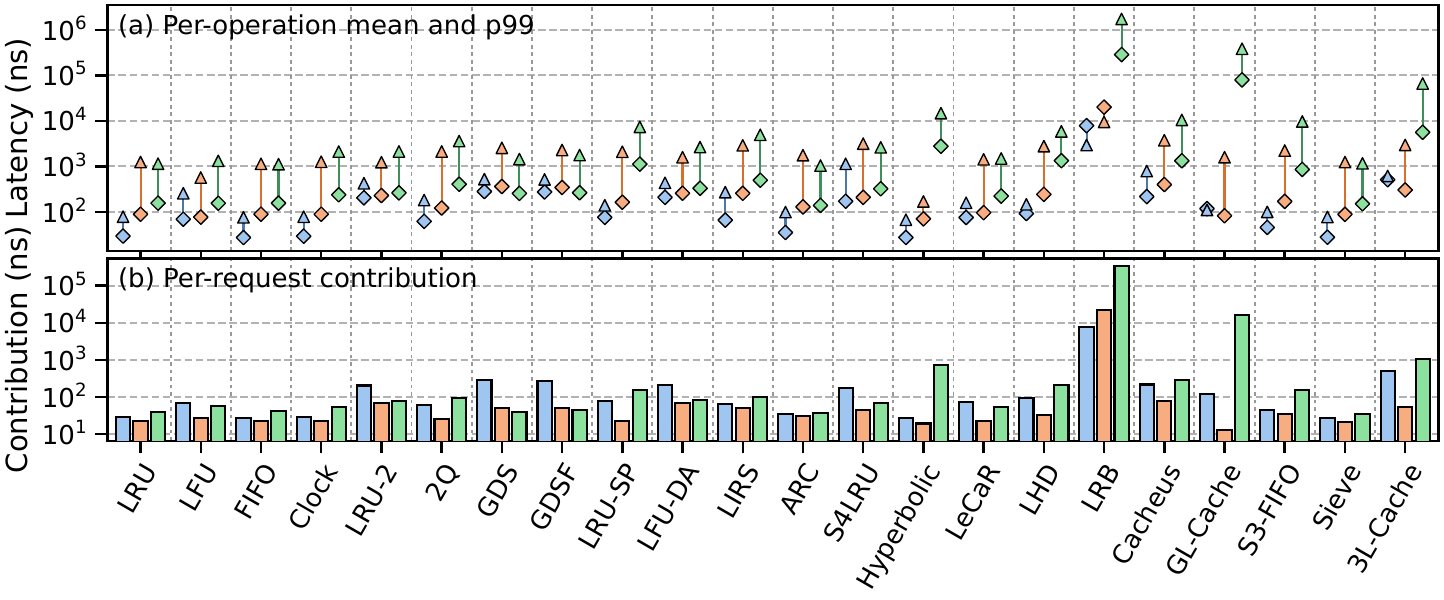}
    
    \vspace{-12pt}
    
    \caption{\Blue{Markers in (a) show the mean and p99 latency of each operation, computed only over requests for which that operation occurs. Bars in (b) show the total time spent in each operation divided by the total number of requests, aggregated using observed operation counts across all workloads and cache sizes. Insert and Evict occur only on misses, whereas Find occurs on every request.}}
    \label{fig:op-overhead}
    \vspace{-7mm}
\end{figure}

\vspace{-2mm}
\subsection{Algorithm Overheads}
\label{subsec:overheads}

An E\&A comparison is incomplete without overhead, because a policy that saves remote downloads but imposes large local cost may not be practical.
We therefore measure both operation latency and metadata footprint.

\textit{$\bullet$ Operation latency.}
For most policies, operation latency remains in the sub-microsecond range; the main exceptions are learning-based methods that incur repeated model maintenance and inference.
We instrument each policy's find, insert, and evict operations using invariant-TSC on a pinned measurement core.
\Blue{Per-operation mean and p99 latencies are computed only over occurrences of each operation.
The amortized per-request contribution sums operation time using the observed Find, Insert, and Evict counts across all workloads and cache sizes, and divides by the total number of requests.
Insert and Evict occur only on misses, whereas Find occurs on every request.}
These measurements provide a controlled baseline for replacement-logic overhead rather than a direct measure of concurrent throughput.
Figure~\ref{fig:op-overhead} shows that, on a per-operation basis, eviction is usually the most expensive operation.
Priority-queue policies such as \GDS{}, \GDSF{}, \LFUDA{}, and \LRUK{} pay $\mathcal{O}(\log N)$ maintenance cost, Hyperbolic pays sampling cost, and \Cacheus{} pays the cost of maintaining two experts and their ghost structures.
\LRB{} and \GLCache{} incur the largest overhead because they perform continuous training and repeated inference: \LRB{}'s per-operation eviction p99 reaches 1.5\,ms, whereas \GLCache{}'s reaches 0.39\,ms.
\ThreeLCache{} reduces this burden by batching predictions and reusing them across multiple evictions.

\begin{table}[t]
\centering
\footnotesize
\caption{Per-object metadata overhead of cache replacement algorithms in bytes.}
\vspace{-10pt}
\label{tab:metadata}
\begin{tabular}{lr@{\hskip 1.5em}lr@{\hskip 1.5em}lr}
\toprule
Algo. & Ovhd. & Algo. & Ovhd. & Algo. & Ovhd. \\
\midrule
\LRU{}       & 16   & \LRUSP{}      & 36   & \LRB{}        & 189  \\
\LFU{}       & 24   & \LFUDA{}      & 28   & \Cacheus{}    & 58   \\
\FIFO{}      & 16   & \LIRS{}       & 18   & \GLCache{}    & 30.4 \\
\Clock{}     & 17   & \ARC{}        & 36   & \SFIFO{}      & 31.4 \\
\LRUTwo{}    & 24   & \SFourLRU{}   & 16   & \Sieve{}      & 17   \\
\TwoQ{}      & 24   & \Hyperbolic{} & 8    & \ThreeLCache{}& 86   \\
\GDS{}       & 28   & \LeCaR{}      & 82   &               &      \\
\GDSF{}      & 36   & \LHD{}        & 33   &               &      \\
\bottomrule
\end{tabular}
\vspace{-15pt}
\end{table}

\textit{$\bullet$ Metadata footprint.}
Metadata overhead is small for all policies, although learned prediction-based methods require the most state.
Table~\ref{tab:metadata} reports the per-object metadata required beyond the shared lookup structure.
Basic list- and queue-based policies require only a few pointers and counters, size-aware policies add size or score information, and adaptive meta-policies add ghost-state and expert-state maintenance.
Learned policies require the most metadata because they track feature histories and model-related state.
Even so, the absolute footprint remains negligible relative to cloud-analytics object sizes.
Using the smallest mean object size in our workloads, 1.49\,MiB in MTD-DS (Table~\ref{tab:uhr-weights}), as a conservative reference, even the heaviest policy's metadata remains only a tiny fraction of cached object size.
In practice, memory overhead is rarely the limiting factor; CPU overhead is the dominant concern.

\vspace{-2mm}
\section{Practical Guidance for Choosing Cache Policies}
\label{sec:practical_guidance}

This section distills Section~\ref{sec:empirical} into a selection map for choosing cache-replacement policies in direct-access cloud OLAP.
Our goal is not to identify a single universally best policy, but to determine which policy families fit which objectives and workload traits. 
Each recommendation below is grounded in the evidence from Section~\ref{sec:empirical}.

\textit{$\bullet$ Choose the objective first.}
Use \OURHRSHORT{} when the deployment goal is to reduce remote-download latency, because cloud misses combine fixed request overhead with size-dependent transfer cost and \OURHRSHORT{} best tracks measured S3 latency. 
Use OHR when request-count efficiency matters most, and BHR when transferred bytes or byte-based cost dominate.

\textit{$\bullet$ When OHR matters, prefer size-aware policies.}
Under variable-sized objects, size normalization is the dominant driver of OHR.
Policies such as \GDS{}, \GDSF{}, \LRUSP{}, and \LHD{} consistently outperform non-size-aware baselines because they retain more objects under the same cache budget.
Notably, this effect is not limited to recent learning-based policies: classical size-aware heuristics remain highly competitive on this objective.

\textit{$\bullet$ When BHR matters, prioritize zero-reuse filtering.}
Strong BHR begins with evicting one-hit objects early.
Under tight cache pressure, policies such as \Cacheus{}, \SFIFO{}, and \LIRS{} perform well because they reduce wasted byte-time on objects that are admitted once and never reused.
As caches grow, filtering remains necessary but allocation quality among reusable objects becomes the main differentiator.

\textit{$\bullet$ Use \Blue{latency-cost} only on top of size-correct scoring.}
A \Blue{latency-cost} term is beneficial only when the underlying score already accounts for object size correctly: \GDS{}-Latency and \GDSF{}-Latency improve over their uniform-cost variants, whereas \LFUDA{}-Latency worsens because it overprotects large objects.

\textit{$\bullet$ Under workload drift, prefer lightweight adaptivity.}
When access patterns shift over time, lightweight online adaptivity is usually more robust than a fixed recency--frequency balance, making \Cacheus{} and \ARC{} strong default choices. ML-based policies are most suitable when the workload is stable enough for learned priorities to remain valid and when the predictive gain justifies the training and inference overhead.

\textit{$\bullet$ Use semantic caching selectively.}
Semantic caching is most attractive when reuse is highly selective, sparse, and region-like.
In our evaluation, \Crystal{} achieves high BHR at small cache sizes in this regime by materializing only qualifying tuples rather than full column chunks.
Outside this regime, well-tuned chunk caching is usually the safer default because partial column overlap limits reuse and OR-cache refresh adapts slowly.

\textit{$\bullet$ Putting the guidance together.}
These results should be read as a selection map rather than as a single ranking.
In practice, first choose the objective (\OURHRSHORT{}, OHR, or BHR), then match the dominant workload trait: variable object sizes call for size-aware policies, many one-hit objects call for strong reuse filtering, drifting phases favor lightweight adaptation, and highly selective predicate reuse favors semantic caching.
Finally, apply operational constraints: when CPU overhead or rapid drift matters, prefer lightweight adaptive policies; when the workload is stable and predictive gains are large, ML-based methods become more attractive.
\Cacheus{} is a strong default when filtering and adaptivity are both important.
Section~\ref{sec:ouralgo} instantiates the design-level recommendations---latency-aware priority, zero-reuse filtering, and online adaptivity---whereas ML-based policies and semantic caching remain deployment-level rules.

\vspace{-2.5mm}
\section{A Reference Instantiation of the Guidance}
\label{sec:ouralgo}

Section~\ref{sec:practical_guidance} distilled our empirical results into practical guidance for direct-access cloud OLAP caching.
To show that the design-level guidance is constructive rather than merely descriptive, we instantiate three recommendations---latency-aware priority, zero-reuse filtering, and online adaptivity---in a compact reference policy, \OURALGO{} (\OURALGOMEANING{}).
\OURALGO{} should be read as a reference instantiation of the empirical guidance rather than the paper's primary contribution.
We present \OURALGO{} as the \OURHRSHORT{}-oriented instance and \OURALGO{}-OHR as a one-line variant that targets OHR directly.

\vspace{-2.5mm}
\subsection{Policy Design}
\label{sec:ouralgo_design}

A useful reference instantiation for this setting should satisfy three empirical requirements: eviction priority should reflect remote-download latency per cached byte; the policy should filter one-hit objects without sacrificing allocation quality among reusable objects; and it should adapt its recency--frequency balance online as workload phases change. Section~\ref{subsec:broad-comparison} showed that \Blue{latency-cost} variants of \GDS{} and \GDSF{} improve over their uniform-cost forms, but static scoring alone does not clearly dominate the strongest baselines. \OURALGO{} therefore combines a size-normalized latency priority with a lightweight adaptive two-expert meta-policy.

We define the latency efficiency of object $o$ as:
\begin{equation}
\label{eq:phi}
\phi(o)=\frac{L(o)}{S(o)}=\frac{\alpha+\beta S(o)}{S(o)}=\frac{\alpha}{S(o)}+\beta
\end{equation}
This quantity is the remote-download latency saved per cached byte under the miss-cost model of Section~\ref{sec:metric-lens}. \OURALGO{} maintains two experts over a shared physical cache. The frequency expert ranks objects by:
\begin{equation}
\label{eq:pr_lfu}
\text{Pr}_{\text{LFU}}(o)=\text{Freq}(o)\cdot \phi(o)
\end{equation}
which mirrors \GDSF{}'s principle of combining reuse with a size-normalized cost term, 
but omits the global inflation factor $L$ to keep the frequency expert purely frequency-driven. Under stationary access patterns, recency carries little additional signal beyond frequency, so $L$ would add noise that weakens zero-reuse filtering, the expert's main role.

The frequency expert serves two complementary purposes. First, it filters zero-reuse objects: low $\text{Freq}(o)$ yields low priority, so one-hit admissions are evicted quickly. Second, to avoid penalizing objects whose first cache residency was cut short by capacity pressure rather than by genuine non-reuse, the expert consults a bounded metadata-only ghost structure, $\mathrm{Ghost}_{\text{Freq}}$, that preserves access counts across transient evictions so that a re-admitted object recovers its previous count rather than restarting from Freq = 1.

\begin{figure}[t]
  \centering
  
  \includegraphics[width=0.68\columnwidth]{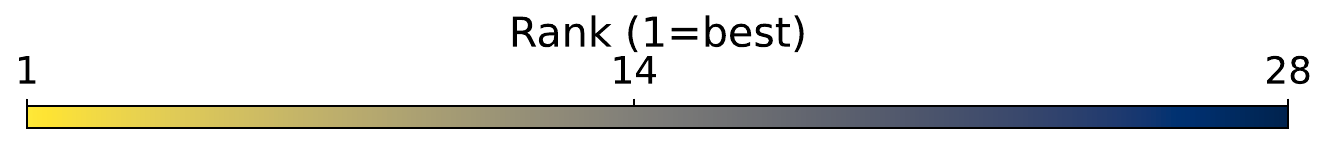}

  \begin{subfigure}[b]{\columnwidth}
     \centering
     \includegraphics[width=0.75\columnwidth]{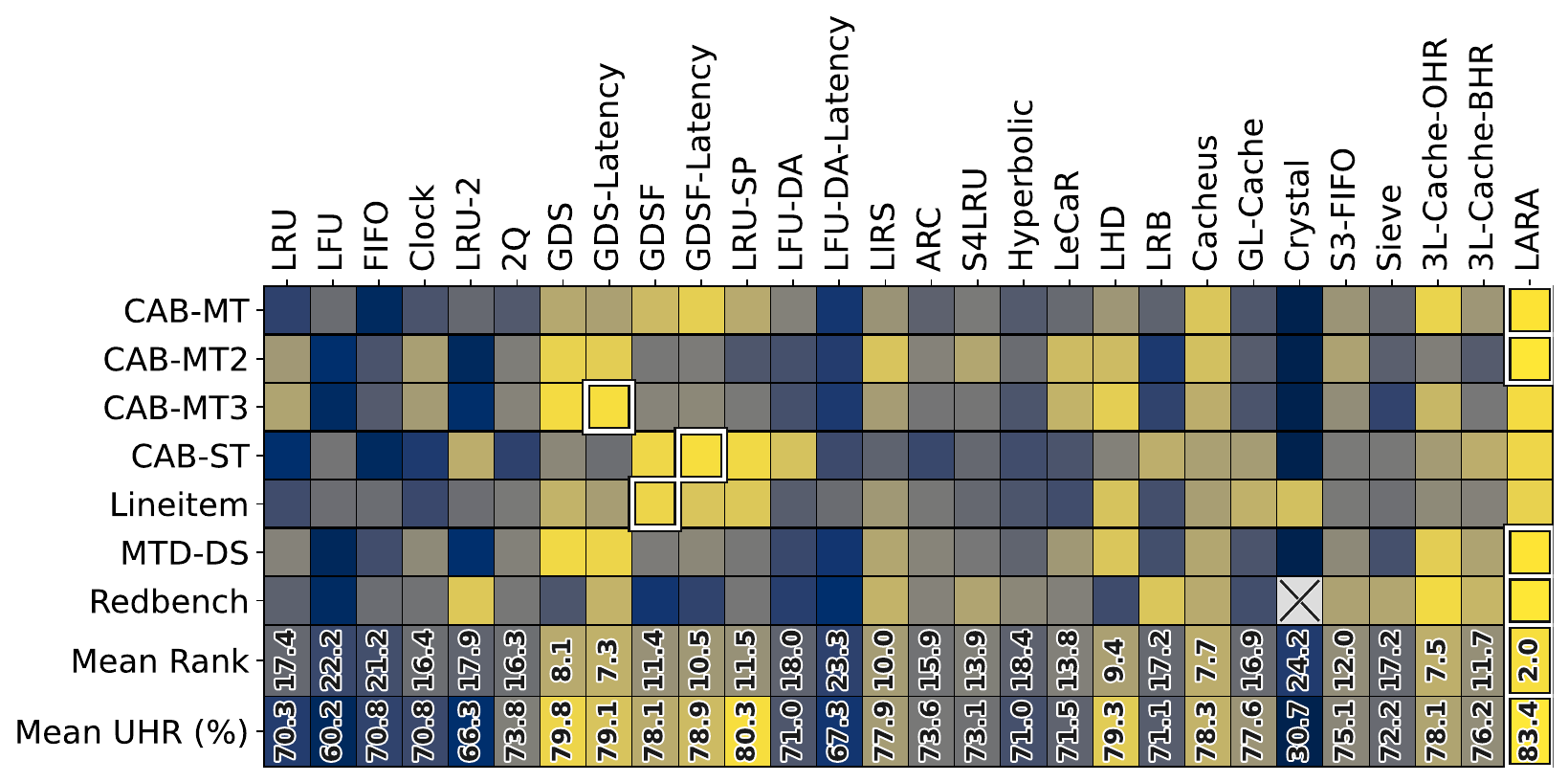}
    \vspace{-4.5pt}
    \caption{\OURHRSHORT{}}
    \label{fig:sim_uhr_laputa}
  \end{subfigure}

  \begin{subfigure}[b]{\columnwidth}
     \centering
     \includegraphics[width=0.75\columnwidth]{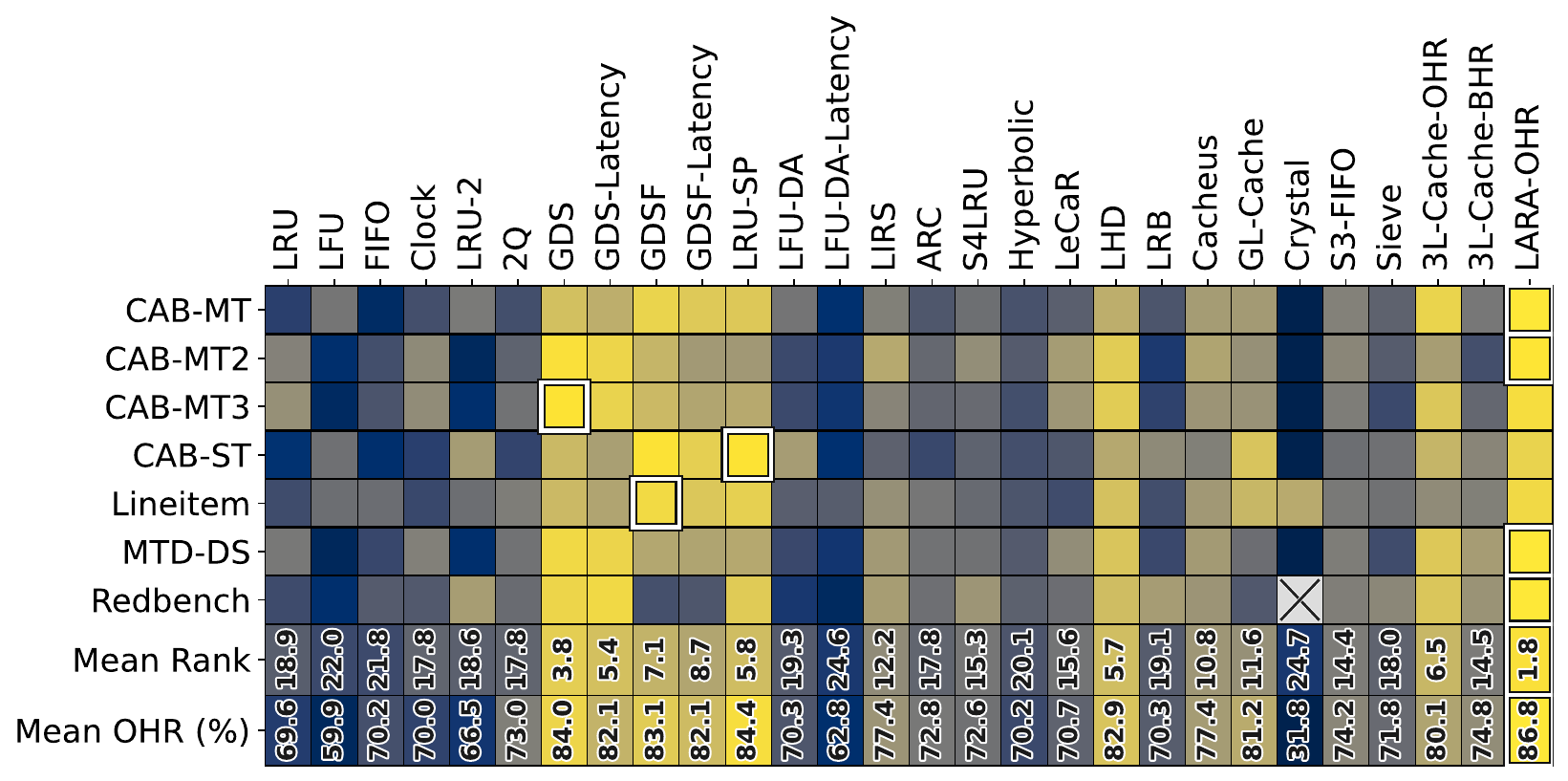}
    \vspace{-4.5pt}
    \caption{OHR}
    \label{fig:sim_ohr_laputa}
  \end{subfigure}

  \vspace{-4mm}
  \caption{Mean \OURHRSHORT{} and OHR rank heatmaps aggregated over ten cache sizes per workload, with \OURALGO{} (or \OURALGO{}-OHR) included. Lower rank is better.}
  \label{fig:simulation_uhr_ohr_laputa}
\end{figure}

The recency expert ranks objects by:
\begin{equation}
\label{eq:pr_lru}
\text{Pr}_{\text{LRU}}(o)=\frac{1}{\text{Age}(o)}\cdot \phi(o)
\end{equation}
where $\text{Age}(o)$ is the logical time since the last access. This lets the policy react quickly to phase shifts while keeping the score aligned with \OURHRSHORT{}.

The two experts are coordinated by a lightweight stochastic meta-policy inspired by \LeCaR{} and \Cacheus{}. It maintains a dynamic weight $W_{\text{LRU}}\in[0,1]$, initialized to $0.5$, which is interpreted as the probability of selecting the recency expert when an eviction is required. When the cache is full, we draw $u\sim \mathrm{Uniform}(0,1)$ and select the recency expert if $u<W_{\text{LRU}}$ and the frequency expert otherwise. To provide feedback, the meta-policy keeps two bounded metadata-only ghost lists, $\mathrm{Ghost}_{\text{LRU}}$ and $\mathrm{Ghost}_{\text{LFU}}$, one per expert. 
Here, $\mathrm{Ghost}_{\text{LRU}}$ and $\mathrm{Ghost}_{\text{LFU}}$ are maintained separately from $\mathrm{Ghost}_{\text{Freq}}$ introduced above: the former two record recent evictions for meta-policy weight updates, whereas $\mathrm{Ghost}_{\text{Freq}}$ is consulted only by the frequency expert to restore prior access counts upon re-admission.
A miss that hits one of these ghosts is treated as evidence that the corresponding expert evicted too aggressively. Let $W_{\text{LFU}}=1-W_{\text{LRU}}$. On a miss for object $o$, we therefore update:
\vspace{-1mm}
\[
(\tilde W_{\text{LRU}},\tilde W_{\text{LFU}})=
\begin{cases}
\big(W_{\text{LRU}}e^{-\eta},\,W_{\text{LFU}}\big), & o\in\mathrm{Ghost}_{\text{LRU}},\\
\big(W_{\text{LRU}},\,W_{\text{LFU}}e^{-\eta}\big), & o\in\mathrm{Ghost}_{\text{LFU}},\\
\big(W_{\text{LRU}},\,W_{\text{LFU}}\big), & \text{otherwise},
\end{cases}
\vspace{-1mm}
\]
followed by normalization,
\[
W_{\text{LRU}}\leftarrow
\frac{\tilde W_{\text{LRU}}}{\tilde W_{\text{LRU}}+\tilde W_{\text{LFU}}},
\qquad
W_{\text{LFU}}\leftarrow 1-W_{\text{LRU}}
\]
This yields lightweight per-request adaptivity without explicit cache partitioning.

The OHR-oriented variant keeps the same architecture but replaces $\phi(o)$ with $1/S(o)$ in both experts, treating all misses as having equal latency cost and directly optimizing request-count efficiency per cached byte.

\vspace{-2.5mm}
\subsection{Evaluation of \OURALGO{}}
\label{sec:laputa_evaluation}

We evaluate \OURALGO{} using the same methodology as in Section~\ref{sec:empirical}. Figure~\ref{fig:simulation_uhr_ohr_laputa} summarizes the rank heatmaps after adding \OURALGO{} and \OURALGO{}-OHR to the benchmark.

\textit{Main result.}
A compact policy built from the three design elements remains competitive under UHR.
In our benchmark, \OURALGO{}  attains the strongest mean UHR rank (2.0) and remains within the top four across all evaluated workloads.
We interpret it as suggesting that the design-level guidance can be combined coherently in a single lightweight policy, rather than that a newly introduced policy should supersede the benchmark’s comparative findings.

\begin{figure}[t]
  \centering
  \includegraphics[width=0.6\columnwidth]{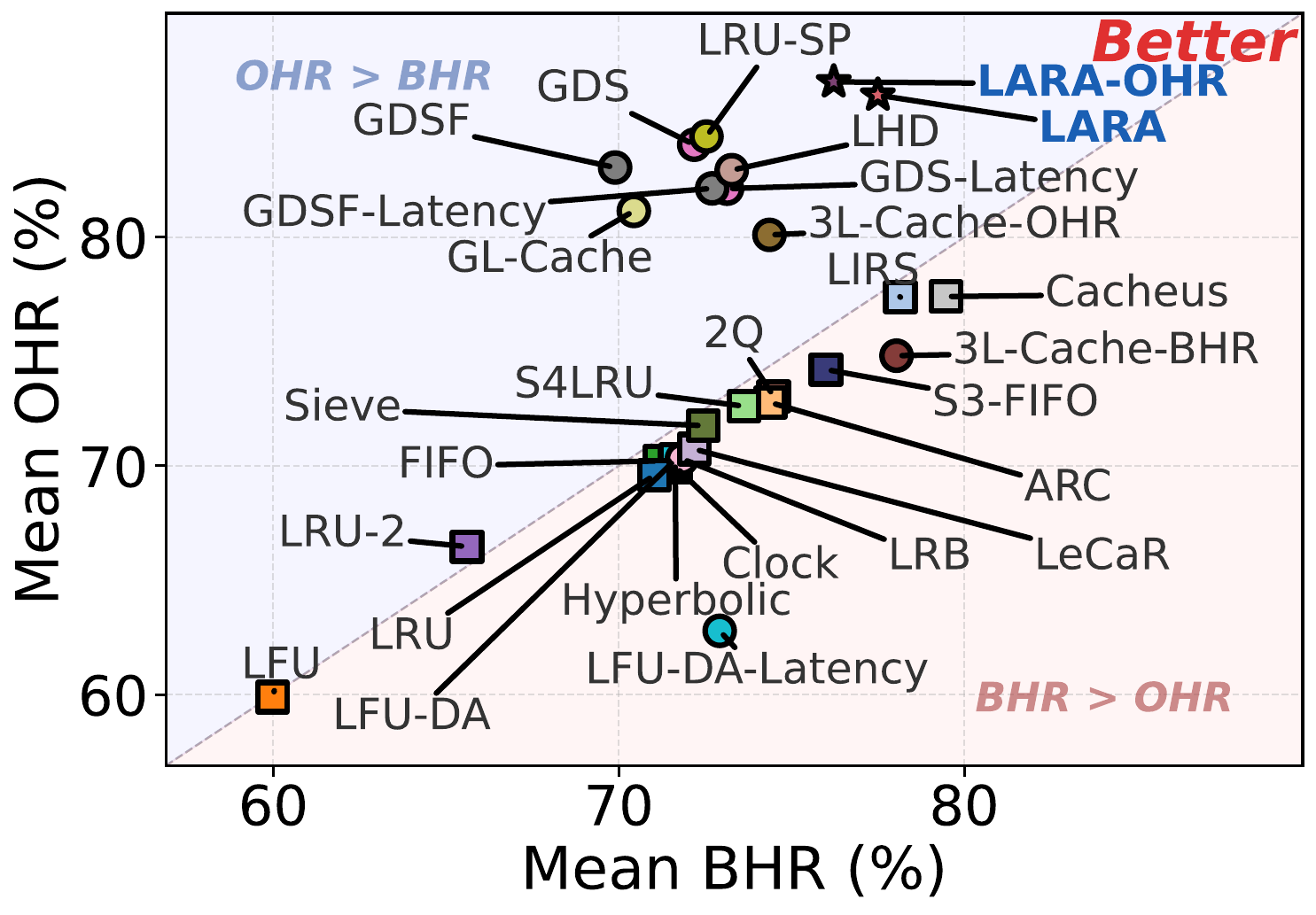}
  \vspace{-3mm}
  \caption{Mean OHR vs.\ mean BHR averaged over all configurations. Higher is better on both axes. Circles denote size-aware algorithms, squares denote non-size-aware algorithms, and stars denote \OURALGO{} and \OURALGO{}-OHR.}
  \vspace{-5mm}
  \label{fig:ohr_vs_bhr}
\end{figure}

\textit{Trade-off and OHR-tuned variant.}
Figure~\ref{fig:ohr_vs_bhr}  places \OURALGO{} near the upper-right frontier of the OHR–BHR plane, alongside strong size-aware and strong filtering-oriented baselines. The point of this figure is descriptive rather than competitive: combining latency-aware priority, zero-reuse filtering, and online adaptivity yields a balanced operating point on the frontier. Replacing $\phi(o)$ with $1/S(o)$ shifts the reference instantiation upward along the OHR axis, as expected from the objective change. The OHR-oriented variant also behaves as expected quantitatively, improving OHR relative to the UHR-oriented reference instantiation.

\textit{Operational overhead.}
Operationally, the reference instantiation remains in the same practical range as strong adaptive baselines. It uses 84 bytes of metadata per object and incurs an amortized per-request cost of 485 ns, so this constructive check does not rely on unusually high local overhead.

\vspace{-3mm}
\section{\revision{Discussion}}
\label{sec:discussion}

\vspace{-1mm}
\subsubsection*{$\bullet$ Beneficiaries and open problems:}

\revision{
Our results provide policy-selection guidance for implementers and operators of direct-access cloud OLAP and lakehouse caches, and give cache researchers an empirical basis, benchmark setup, artifacts, and UHR validation protocol for further study.
The study provides guidance for cache replacement rather than the complete cache-design space. Admission, concurrent eviction, and pushdown-aware caching remain open.
}

\vspace{-3mm}
\subsubsection*{$\bullet$ Cache granularity:}
\revision{Object-size variability, reuse, zero-reuse filtering, and workload drift remain relevant at coarser cache units such as files.
However, changing the cache unit changes the object-size distribution and request stream. Exact policy rankings may therefore differ.}

\vspace{-3mm}
\subsubsection*{$\bullet$ Remote predicate pushdown:}
\revision{Remote predicate pushdown changes the caching problem because the request stream and effective miss cost become predicate-dependent. A policy must also consider expected pushdown benefit.
Pushdown-aware caching may change exact rankings, although object-size variability, zero-reuse filtering, and workload drift still affect replacement decisions.}

\vspace{-3mm}
\subsubsection*{$\bullet$ Broader applicability:}
\revision{These insights may also apply to other systems that cache variable-sized objects with nonuniform miss costs, including serverless and dataframe analytics over object stores and CDN object caching, although their applicability needs to be evaluated for each workload.}

\vspace{-3mm}
\section{Conclusion}
\label{sec:conclu}
We presented a broad experimental comparison of cache replacement policies for direct-access cloud OLAP under compute-storage disaggregation and derived practitioner-oriented guidance from it.
To evaluate remote-I/O benefit on a common basis, we formulated Unified Hit Ratio (UHR), a latency-oriented metric that measures the normalized fraction of remote-download latency saved by caching, and validated it against live object-store replay.
Across 22 representative algorithms, their \Blue{latency-cost} and objective variants, and one semantic-caching baseline, our results show that the design emphasis shifts with the objective: size-awareness drives OHR, aggressive zero-reuse filtering drives BHR, and the two meet on a common UHR frontier under the cloud miss-cost model, with lightweight online adaptivity outperforming retraining-heavy learned methods under workload drift.
\OURALGO{} is included only as a compact reference instantiation of this design-level guidance---latency-aware priority, zero-reuse filtering, and online adaptivity. The paper's primary contribution remains the broad comparative analysis, the UHR metric, and the resulting practitioner-oriented guidance.

\begin{acks}
This work was partly supported by the National Research Foundation of Korea(NRF) grant funded by the Korea government(MSIT) (RS-2025-00517736, 85\%) and the Institute of Information \& Communications Technology Planning \& Evaluation (IITP) grant funded by the Korea government(MSIT) (RS-2026-25529760, ITRC, 5\%; No. RS-2024-00509258, Global AI Frontier Lab, 5\%; IITP-2026-RS-2026-25616370, AI Star Fellowship Support Program, 5\%).
\end{acks}

{
\small
\bibliographystyle{ACM-Reference-Format}
\bibliography{refs}
}

\end{document}